\documentclass[11pt, oneside]{article}   	
\usepackage{geometry}                		
\pdfoutput=1
\usepackage[parfill]{parskip}    		
\usepackage{graphicx}				
\usepackage{amssymb}
\usepackage{amsmath}
\usepackage{bm}

\title{Brief Article}
\author{The Author}

\def\blp{\bigg(}
\def\brp{\bigg)}

\def\l{\lambda}
\def\la{\langle}
\def\ra{\rangle}

\def\j{\varphi}
\def\Mv{{\mathbf{M}}}

\def\Bv{{\mathbf{B}}}

\def\xv{{\mathbf{x}}}
\def\yv{{\mathbf{y}}}
\def\zv{{\mathbf{z}}}

\renewcommand\phi{\varphi}

\def\kv{{\mathbf{k}}}

\def\l{\lambda}

\def\Lv{{\mathbf{L}}}

\def\Mv{{\mathbf{M}}}

\title{Generalized Erd\H{o}s Numbers for network analysis}

\author{
Greg Morrison$^{1,2}$, Levi Dudte$^1$, and L. Mahadevan$^1$\\
1:  School of Engineering and Applied Sciences,\\Harvard University, Cambridge MA 02138\\
2:  Laboratory for the Analysis of Complex Economic Systems,\\ IMT Institute for Advanced Studies, Lucca Italy 55100
}

\begin{document}
\maketitle

\maketitle

\begin{abstract} 
In this paper we consider the concept of `closeness' between nodes in a weighted network that can be defined topologically even in the absence of a metric. The Generalized Erd\H{o}s Numbers (GENs) \cite{MorrisonEPL11} satisfy a number of desirable properties as a measure of topological closeness when nodes share a finite resource between nodes as they are real-valued and non-local, and can be used to create an asymmetric matrix of connectivities. We show that they can be used to define a personalized measure of the importance of nodes in a network with a natural interpretation that leads to a new global measure of centrality and is highly correlated with Page Rank.  The relative asymmetry of the GENs (due to their non-metric definition) is linked also to the asymmetry in the mean first passage time between nodes in a random walk, and we use a linearized form of the GENs to develop a continuum model for `closeness' in spatial networks. As an example of their practicality, we deploy them to characterize the structure of static networks and show how it relates to dynamics on networks in such situations as the spread of an epidemic.
\end{abstract}

The study of complex networks has increased enormously in recent years due to their applicability to a wide range of physical\cite{BassettPRE12,BarthelemyPhysRep2011}, biological\cite{BascompteScience06}, epidemiological\cite{NewmanPRE02,BalcanPNAS09}, and sociological\cite{KeelingThPopBiol05} systems.  Two basic goals in this regard are to understand and quantify the structure of the network to better characterize the relationship between the interacting members of the network (the nodes), while also characterizing the dynamical processes on the network\cite{KeelingThPopBiol05} that may shed light on the processes by which they form \cite{BarabasiScience99}.    

Understanding the topological properties of the network on both a global and local level can be useful in approaching both of these goals.  Global properties of interest may include simple measures of the distribution of node properties, such as the degree distribution, $P_d(k)$, with $k$ the number of edges incident upon a node; strength distribution, $P_s(s)$, with $s$ the total weight incident upon a node; or distribution of clustering coefficients, $P_c(c)$, with $c$ the fraction of triplets of connected nodes that are closed\cite{BarratPNAS04,NetworkBook}.  Community structure in the network\cite{LFR,MorrisonPLoS12,GirvanPNAS02}, which partitions the network into densely connected sub-networks with more links within communities than between communities, has been extensively studied and may provide more detailed information about the relationship between nodes than simple distributions.  Community structure in networks can indicate the existence of underlying similarities between nodes in the network, and may have a great impact on dynamical processes occurring on the network (such as a random walk\cite{NohPRL04,KleinJMathChem93,NewmanSocNet05} or epidemic spreading\cite{NewmanPRE02,SatorrasPRL01,BallMathBio2008}), and can influence the material properties of granular systems\cite{BassettPRE12}.  

While global properties of networks can be used to assess the attributes of the nodes on an aggregate level, it is also of great interest to understand the topological properties of nodes on an individual, local level as well.  Node centrality is the classic example of a topological measure of an individual node, which assesses the `importance' of a node in a variety of contexts.  The most basic measure of a node's centrality is simply related to it's degree, a property of the node that is based solely on the local topology of its connectivity.  The centrality of individual nodes can also be measured incorporating the global topology of the network in a variety of ways, including PageRank\cite{FranceschetComACM11}, betweenness\cite{NewmanSocNet05}, or random walk\cite{NohPRL04} centralities.  Each of these measures reduces the global properties of the network into a individualized local measure of importance, permitting a rank-ordering of their importance in the network\cite{GhoshalNatCom11,BlummPRL12}.



\begin{figure}
\begin{center}
\includegraphics[width=.4\textwidth]{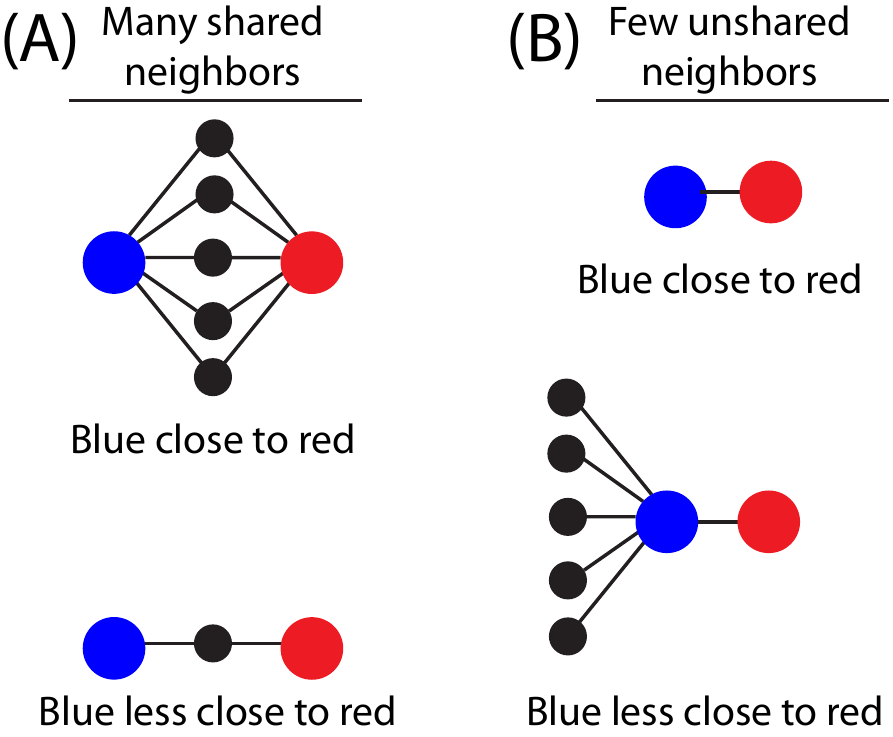}
\caption{Two competing requirements for global `closeness' in a network with shared resources.  In (A), many short paths between nodes increase the closeness between them.  This is similar to the resistance distance between nodes:  additional parallel paths between them reduce their resistance distance.  In (B), the finite resources of the high-degree blue node suggest it should be less close to the red node than for the lower-degree blue node above, as resources are shared with the other neighbors as well.  This is similar to the transition probability from the blue node in a random walk:  the more connections the blue node has, the lower probability of visiting the red node.\label{GENsSchem.fig}
}
\end{center}
\end{figure}


In many contexts\cite{NewmanPRE01,ZhouPRE07}, not all members of the network will necessarily agree on the importance of the same node:  nodes that have a direct connection between them will be more important to each other than distant nodes in the network.  Nodes that are central to the network as a whole may have very low importance from the perspective of sub-networks.  The universality of importance is further complicated by the fact that we may expect the influence between a pair of nodes to be asymmetric even if they are directly connected\cite{ZhouPRE07} (the importance assigned by an important node towards an unimportant one is not necessarily the same as the importance assigned in the opposite direction), which may have important consequences in real-world systems\cite{BascompteScience06}.  The determination of a personalized measure of node importance that incorporates the global topology in an asymmetric measure is therefore an important but non-trivial problem.

In this paper, we use the recently developed Generalized Erd\H{o}s Numbers\cite{MorrisonEPL11,MorrisonPLoS12} (GENs), which provide a nonlinear asymmetric measure of the pairwise `closeness' between nodes based on the network's global topology, as a proxy for the importance one node feels towards another.  We show that this measure of closeness can be linked to the dynamics of random walks and spreading of epidemics on networks.  Using the GENs, we develop a linearized importance that provides an intuitive measure of the (asymmetric) importance one node feels towards another based on the global network topology.  
This linearized importance is considered in a continuum approximation for a spatial network with homogeneous distance-dependent links between material points on a sphere which is coupled to long-range distance-independent connection strengths, and is shown to reduce to an inhomogeneous Fredholm equation with a kernel depending on the topology of the spatially-independent portion of the network.  Finally, we show that a global measure of centrality defined using the GENs is consistent with other centrality measures, and that the asymmetry in the GENs is consistent with the asymmetry of the mean first passage time (MFPT) between nodes in a random walk.  
The flexibility of the GENs in understanding the topology of complex networks shows their utility in a wide range of problems.

\section{The Generalized Erd\H{o}s Numbers\label{GeneralGENsSec}
}

When nodes represent objects in a physical space\cite{WirelessSensor,GeographicNetwork,KurantPRL05,BarthelemyPhysRep2011,MontisEnvPlan07}, the distance between nodes, $D_{ij}$, is a naturally defined (metric) measure of closeness between the objects.  Due to the generality of networks (where nodes and edges abstractly represent `objects' and `interactions,' respectively), there can be no guarantee of a naturally defined distance metric\cite{BoccalettiPhysRep11,BarthelemyPhysRep2011}, and, in some cases, the network topology itself must define a measure of closeness ($\Delta_{ij}$) based solely on the matrix of weights between nodes $i$ and $j$, $w_{ij}$ (with an undirected network where $w_{ij}=w_{ji}$ is assumed throughout this paper).  The closeness, $\Delta_{ij}$, will be small for nodes that are close to one another and large for distant nodes, with a simple and common choice being $\Delta_{ij}=w_{ij}^{-1}$ (so strongly connected nodes are `close', and disconnected nodes are `far').  Alternatively, in an unweighted network, the length of the shortest path between a pair of nodes is a natural definition\cite{BoccalettiPhysRep11,RivesPNAS03} and is the basis for the classic Erd\H{o}s numbers in the context of an unweighted collaboration network\cite{DeCastroMathIntel99}.  Improvements on this measure which incorporate the effect of multiple paths between nodes (see Fig. \ref{GENsSchem.fig}(A) for a schematic diagram) include the resistance distance\cite{KleinJMathChem93,BabicIntJQantChem02}, self consistent similarity measures\cite{LeichtPRE06}, and communicability \cite{EstradaPRE08}, to name only a few.  An additional approach to defining similarity between nodes is found by positing a multidimensional `latent space' of node properties\cite{HoffJasa02}, with the assumption that nodes that are close in the latent space are likely to be connected in the network and each node's position in the space inferred from the observed connectivity.  Each of these methods incorporates the global topology of the network into a symmetric measure of closeness between pairs of nodes ($\Delta_{ij}=\Delta_{ji})$.

Finite resources are shared in some networks, with examples including collaboration on networks (where time with one collaborator reduces the available time for others), multi-core processor components\cite{multicore} (where finite memory or other hardware must be shared), and random walks (where the walker can only move to a single neighbor at a time with a transition probability $P_{i\to j}=w_{ij}/W_i$ with $W_i=\sum_k w_{ik}$ the total strength of the node $i$).  In the context of these networks of limited resources, closeness measures such as resistance distance may be undesirable\cite{ZhouPRE07}, because the addition of a new edge in the network should be detrimental to some nodes (those who receive less of the finite resource due to the new edge) and beneficial to others (those who receive more due to the edge).  For closeness measures based on the direct weight between nodes (where the `closeness' between $i$ and $j$ is often taken to be $w_{ij}^{-1}$) or resistance distance between nodes, it is straightforward to see that the newly measured closeness between nodes $i$ and $j$ $\Delta_{ij}^{(new)}\le \Delta_{ij}^{(old)}$ for all pairs, i.e. the addition of an edge can never cause nodes to feel less close to one another.  This is not sensible in the context of nodes that share a finite resource with its neighbors, as shown in Fig. \ref{GENsSchem.fig}(B):  if a node $i$ has many neighbors, each receives less of the resource than if $i$ had few neighbors.  A quantity such as the transition probability in a random walk, $P_{i \to j}$, is asymmetric and ensures that nodes are closer if they have few neighbors, pictured in Fig. \ref{GENsSchem.fig}(B) (so a walker is more likely to pass between them than if they had many connections).  However, it is not a global measure of closeness because the transition probability incorporates only the nearest neighbor connections between nodes.  It is useful to develop a measure of closeness that incorporates these two (sometimes contradictory) aspects depicted in Fig. \ref{GENsSchem.fig}:  nodes feel close if there are many paths between them, but popular nodes are less close to their neighbors than unpopular nodes.

We have recently shown\cite{MorrisonEPL11} that the Generalized Erd\H{o}s Numbers ($E_{ij}$, or GENs), describing the closeness node $j$ feels towards node $i$, satisfy the expected properties for the sharing of finite resources described in Fig. \ref{GENsSchem.fig}.  The GENs are defined as
\begin{eqnarray}
\frac{W_j}{E_{ij}}=w_{ij}^2+\sum_{\substack{l\in C_j\\l\ne i}}\frac{w_{jl}}{E_{il}+w_{jl}^{-1}}\qquad\qquad E_{ii}\equiv 0,\label{GENdef}
\end{eqnarray}
where $C_j$ is the set of nodes directly connected to $j$. This form is chosen such that the node $i$ is as close as possible to itself and that if $j$ is connected to only one node $k$, $j$'s closeness to $i$ satisfies $E_{ij}\equiv E_{ik}+w_{jk}^{-1}$.  If there are multiple paths between nodes, the closeness $j$ feels to $i$ is strengthened if there is a direct connection between them but also includes a contribution from all other neighbors of $j$ weighted by their connection strength.  By choosing a harmonic mean for the form of the contribution, we bias our measure of closeness towards neighbors that themselves feel close to $i$.  The GENs are defined using the global topology of the network, and $E_{ij}$ is finite even if $i$ and $j$ share no neighbors (as may not be the case for more local measures of closeness \cite{ZhouPRE07}).

This definition of the GENs in Eq. \ref{GENdef} is nonlinear, and the exact values of $E_{ij}$ for complex networks are not easily determined analytically.  $E_{ij}$ can be computed numerically in an iterative fashion\cite{MorrisonEPL11}, with $E_{ij}\equiv E_{ij}^{(\infty)}$ and the recursive definition $W_j/E_{ij}^{(t+1)}=\sum_l w_{jl}/[E_{il}^{(t)}+w_{jl}^{-1}]$ (with the constraint that $E_{ii}^{(t)}=0$ continually enforced).  In this paper, the iteration is halted when $\max_{ij}|E_{ij}^{(t+1)}-E_{ij}^{(t)}|\le \epsilon=0.005$.  The method also requires an initial guess, $E_{ij}^{(0)}$, with the simple initial guess throughout this paper that $E_{ij}^{(0)}=1$ (the method is robust to variations in this initial value). 

Eq. \ref{GENdef} is only one way of satisfying the expectations shown in Fig. \ref{GENsSchem.fig}, and there is a great deal of functional freedom in satisfying these constraints.  For example, any measure $E^{(g)}_{ij}$ of the form $W_j g(E^{(g)}_{ij})=\sum_k w_{ij}g(E^{(g)}_{ij}+w_{ij}^{-1})$ will satisfy the desired behavior depicted in Fig. \ref{GENsSchem.fig} for a monotonically decreasing $g(x)$, with $g(x)=x^{-1}$ in the definition of Eq. \ref{GENdef}.  Another alternate definition will satisfy a triangle inequality at the cost of additional computational complexity by replacing the direct connection strength, $w_{kj}$, with the closeness, $E_{kj}$, in the denominator:  $W_j/{\tilde{E}}_{ij}=w_{ij}^2+\sum_{k\ne i} w_{ij}/({\tilde{E}}_{ik}+{\tilde{E}}_{kj})$.  While these alternate definitions may be of interest in certain contexts, we continue to use Eq. \ref{GENdef} throughout this paper, due to its simplicity and previously demonstrated successes in prediction algorithms\cite{MorrisonEPL11} and community detection methods\cite{MorrisonPLoS12}.  Variations in the definition of $E_{ij}$ will certainly change the numerical values of the closeness, but the qualitative behavior of the closeness between nodes is expected to be robust to perturbations of the definition of the GENs.

\section{The GENs in homogeneous networks}
\label{initalValueSec}

While Eq. \ref{GENdef} is not exactly solvable for all but the simplest of network topologies, the general properties of the GENs can be explored for sufficiently homogeneous networks.  The unweighted Erd\H{o}s-R\'enyi (ER) networks have a degree distribution sharply peaked about the average ($k_i\approx \la k\ra$, where $k_i$ is the degree of the node $i$ in an unweighted network), and we expect the closeness between nodes will still be broadly distributed due to the complex network topology.  The average closeness between nodes can be derived by assuming that $E_{ij}=E_c$ (the `typical connected' closeness) if $i$ and $j$ are connected, and the `typical disconnected' closeness, $E_{ij}=E_d$, if they are not directly connected.  In a unweighted regular network, with all nodes having the same degree $k_i=k$, it is possible to examine the average closeness between connected and disconnected nodes using the GENs.  For homogeneous degree distributions such as the ER networks, we expect an approximation $k_i\approx \la k\ra$ to be reasonable, with fluctuations in the degree expected to have a relatively minor impact, particularly for high mean degree.  For these homogeneous networks, we assume that nodes that are directly connected feel a typical closeness $E_c$ between each other, and another closeness $E_d\ge E_c$ to nodes they are not.  If $i$ and $j$ are directly connected they have on average $(k-1)^2/(N-2)$ neighbors in common (since both have exactly $k$ edges, one of which connects to the other), and they have $k^2/(N-2)$ neighbors in common if they are not connected.  We must split the sum in Eq. \ref{GENdef} into two parts:  a sum over nodes neighboring both $i$ and $j$, and a sum over nodes only connected to $j$.  This gives the approximate equations for an unweighted network of constant degree
\begin{eqnarray}
\frac{k}{E_c}&\approx& 1+\frac{(k-1)^2}{N-2}\ \frac{1}{E_c+1}+\blp k-1-\frac{(k-1)^2}{N-2}\brp\frac{1}{E_d+1}\label{InitialPredictEq1}
\end{eqnarray}
and
\begin{eqnarray}
\frac{k}{E_d}&\approx& \frac{k^2}{N-2}\ \frac{1}{E_c+1}+\blp k-\frac{k^2}{N-2}\brp\frac{1}{E_d+1}\label{InitialPredictEq2}.
\end{eqnarray}
It is possible to solve $E_c$ exactly in terms of $k$, $N$, and the unknown $E_d$, with 
\begin{eqnarray}
E_c=\frac{2+kE_d^2-N}{-2+k E_d+N}.
\end{eqnarray}
 An exact solution for $E_d$ is unwieldy, but it is possible to find a solution for $E_d$ asymptotically for large $N$ and we find that 
\begin{eqnarray}
E_d\sim \blp\frac{(k+1)N}{k}\brp^{\frac{1}{2}}+\frac{k}{k+1}+O(N^{-1/2}).\label{InitErdosDisconnected}
\end{eqnarray}
Comparing this expression to the numerical solution of the equation shows less than 1\% deviation for $N\sim 1000$ and $k\lesssim 300$, suggesting the truncation to terms of order $O(N^0)$ is sufficient for large $N$ over a wide range of $k$.  A good approximation for $E_c$ can be found by setting $k=\kappa N$ and taking the limit of $\kappa \to 0$.  We find
\begin{eqnarray}
E_c\approx \frac{k+2\sqrt{\frac{k^{3}}{N(k+1)}}}{1+\sqrt{\frac{k(k+1)}{N}}}\sim k+O(k^2N^{-\frac{1}{2}})\label{InitErdos}.
\end{eqnarray}
where latter is the scaling for sufficiently large $N$.  Even for $N\sim 1000$, higher order terms can contribute in the series for only moderate values of $k$, and the full expression is required to obtain an accurate estimate.

%

\begin{figure}[tbp]
\begin{center}
\includegraphics[width=.8\textwidth]{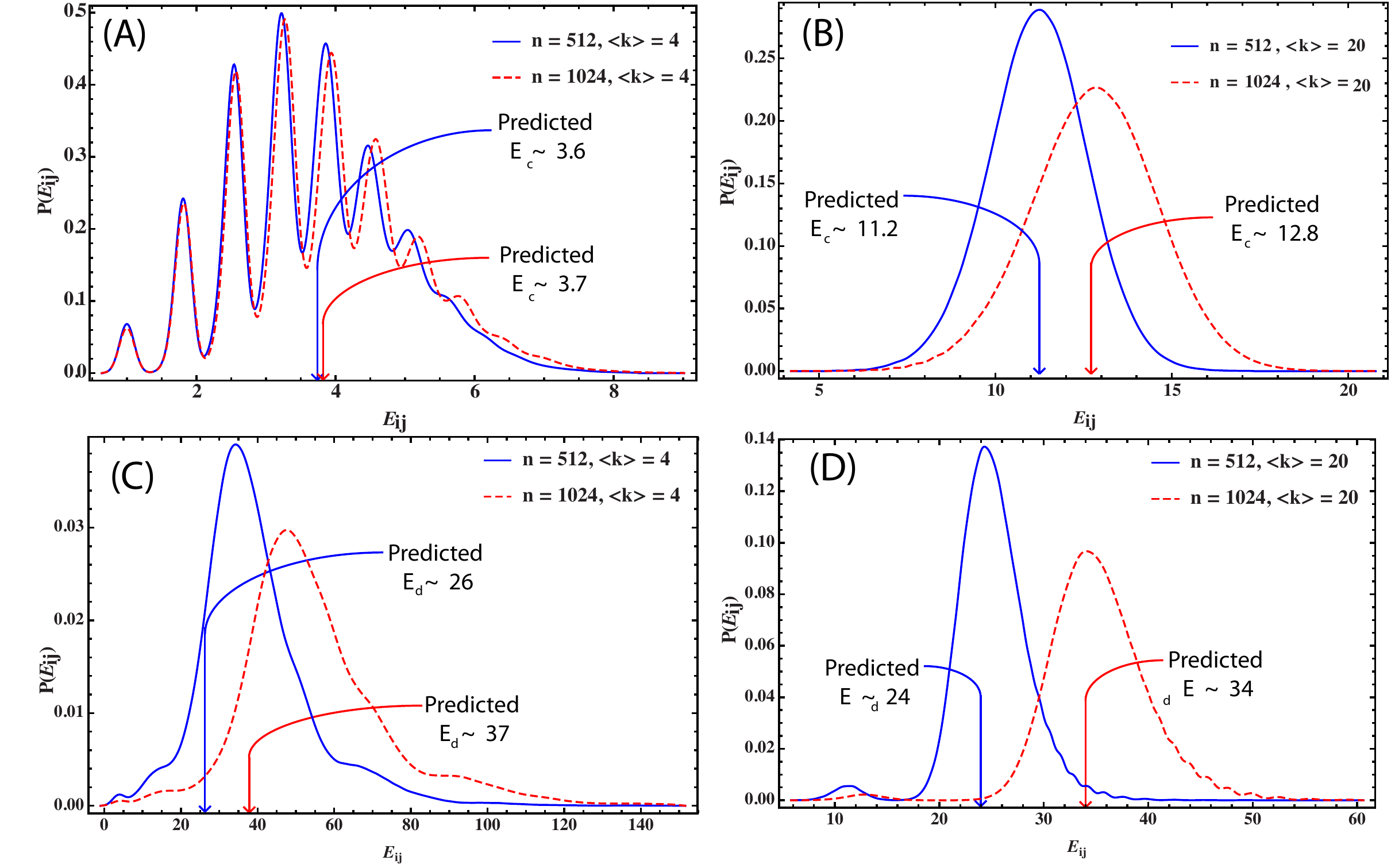}
\caption{The distribution of $E_{ij}$ split into cases where $i$ and $j$ are directly connected in (A,B) and not directly connected in (C,D) for Erd\H{o}s-R\'enyi networks with $N=512$ or 1024 and $\la k\ra=4$ (A,C) or $\la k\ra=20$ (B,D).  Note the changing axes in all figures.  The predicted average of $E_c$ and $E_d$ are marked using the same color schemes as in the figures (where the predictions include the higher-order terms found in Appendix A).  For $\la k\ra=20$, there is excellent agreement between the theoretical and simulated averages.  For $\la k\ra=4$, the GENs are far more heterogeneous than can be captured using the simple model in Eqs. \ref{InitErdosDisconnected} and \ref{InitErdos}, and the theoretical predictions do not agree well with the observed behavior for both connected and disconnected nodes.}
\label{InitialValue.fig}
\end{center}
\rule{\textwidth}{.2pt}
\end{figure}

In Fig. \ref{InitialValue.fig}, we see the distribution of the GENs for Erd\H{o}s-R\'enyi networks with varying $N$ and $\la k\ra$.  In (A-B) we see that changing $\la k\ra$ radically alters the mean values of $E_{ij}$ as well as the shape of the distributions, while changing $N$ only marginally affects the distribution of the connected GENs.  For $\la k\ra=4$ the distribution of $E_{ij}$ exhibits multiple peaks in Fig. \ref{InitialValue.fig}(A), with each local maximum corresponding to a different degree of the node $j$ and with the width of the distribution about the peak coming from differing degrees of the node $i$.  Such heterogeneity is less apparent for high-degree nodes (Fig. \ref{InitialValue.fig}(B)), where fluctuations in the degree of $i$ or $j$ have less of an impact on the GENs, and the distributions are unimodal.  For disconnected nodes, the distributions have a single dominant peak (Fig. \ref{InitialValue.fig}(C-D)), and the location of the peaks are well predicted for $\la k\ra=20$.  Due to the heterogeneity and significance of degree fluctuations for the smaller $\la k\ra=4$, there are large differences between the predicted and observed averages.  

\begin{figure}[tbp]
\begin{center}
\includegraphics[width=.9\textwidth]{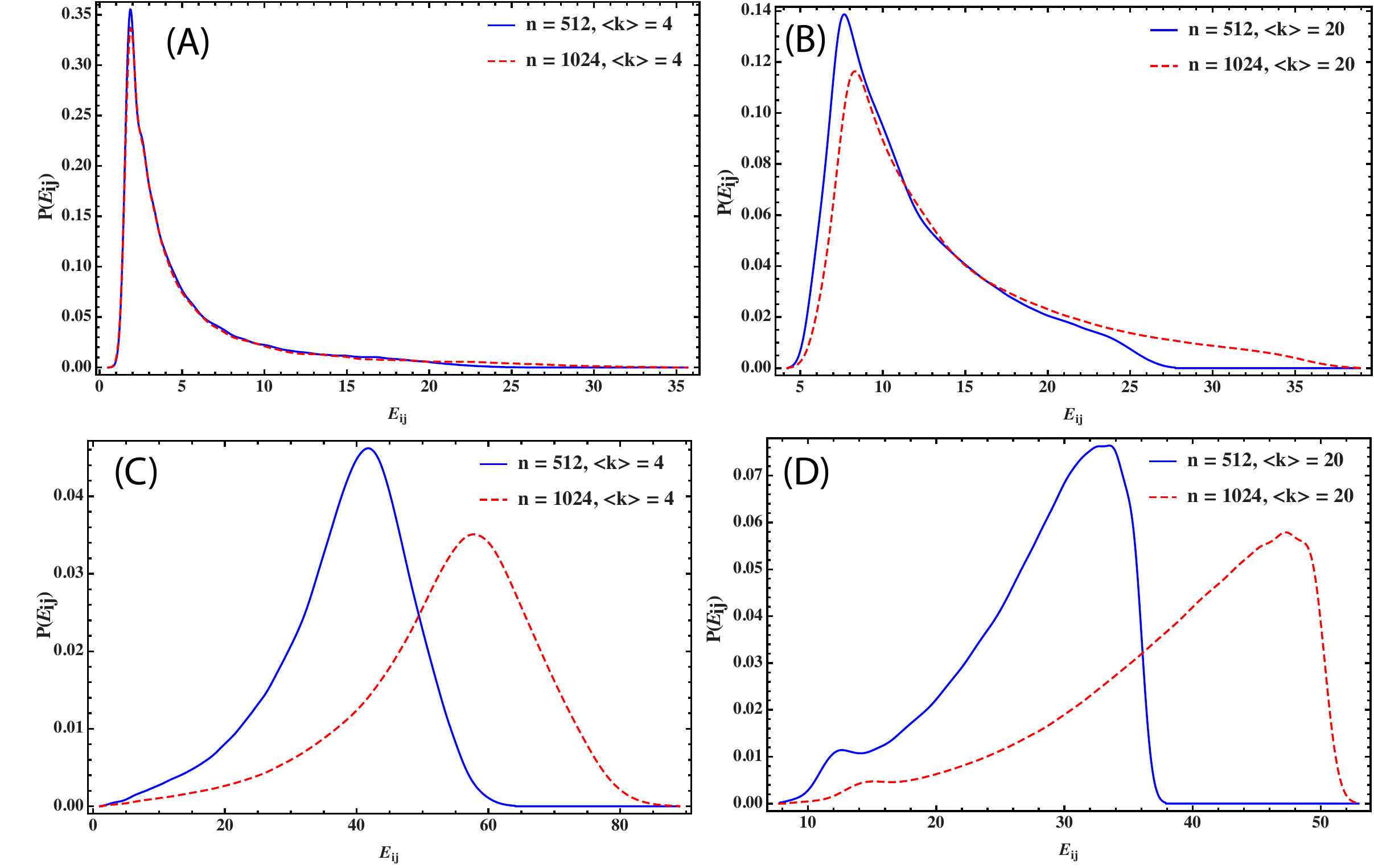}
\caption{
Distributions of the GENs for the Barabasi-Albert networks for nodes that do share a direct connection (A,B) and do not share a connection (C,D) for $\la k\ra=4$ (A,C) and $\la k\ra=20$ (B,D) and with $N=512$ (blue solid lines) or 1024 (red dashed lines).  The distributions are far smoother than those seen in Fig. \ref{InitialValue.fig} for small $\la k\ra$ due to the more heterogeneous degree distribution of the BA networks.  The behavior is consistent with the ER networks:  the mean values depend strongly on $\la k\ra$ and weakly on $N$ if the nodes share a direct connection, while the opposite is true if the nodes are not neighbors.  Because the fat tail of the degree distribution (with $P(k)\sim k^{-3}$) provides a broader distribution to the connected GENs distribution than in the ER case (Fig. \ref{InitialValue.fig}), the mean values are further from the peaks of the distribution and are not indicated in the figure.
}
\label{BarabasiScaling.fig}
\end{center}
\end{figure}

\begin{figure}[tbp]
\begin{center}
\includegraphics[width=.7\textwidth]{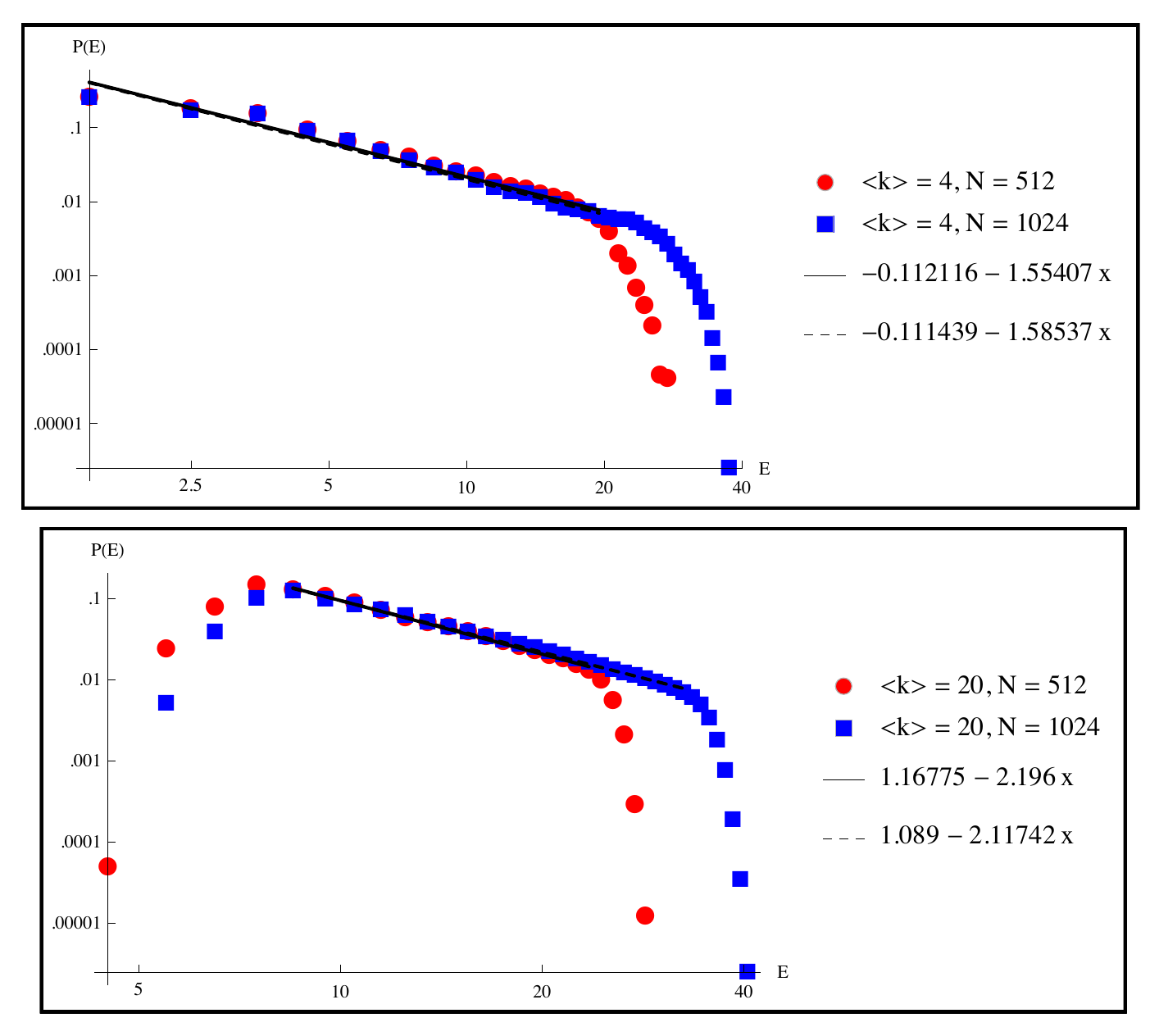}
\caption{
Fitting the heavy tail of the distribution of the nearest-neighbor GENs for the Barabasi-Albert networks in Fig. \ref{BarabasiScaling.fig} (A-B).  Over a wide range of values, there is an approximate power law decay which is very weakly dependent on $N$ ($N=512$ in the red circles and $N=1024$ in the blue squares) but does depend on the average connectivity ($\la k\ra=4$ shown above and $\la k\ra=20$ below).  This is consistent with the weak $N$ dependence seen for the Erd\H{o}s-R\'enyi networks in Eqs. \ref{InitErdosDisconnected} and \ref{InitErdos}.}
\label{DistributionScaleFreeFit.fig}
\end{center}
\end{figure}

In contrast to the homogeneous degree distribution of the Erd\H{o}s-R\'enyi random network model, Barabasi-Albert (BA) networks\cite{BarabasiScience99} have a scale-free, heterogeneous degree distribution, and Fig. \ref{BarabasiScaling.fig} shows that the distribution for the GENs for BA networks are likewise heterogeneous for directly connected nodes.  The distribution for the GENs between nodes that share an edge (shown in Fig. \ref{BarabasiScaling.fig}A-B) appear to have a relatively fat tail and approximately satisfy $Pr(E_{ij}=E)\sim E^{-\lambda}$ for nodes that share a direct connection, with an empirically determined scaling exponent near 1.5 for $\la k\ra$=4 and around 2.1-2.2 for $\la k \ra=20$ (shown in Fig. \ref{DistributionScaleFreeFit.fig}).  This is in comparison to the fat-tailed degree distribution with the $P(k)\sim k^{-3}$ scaling of the BA networks for both values of $\la k\ra$.  Interestingly, the distribution of the GENs for disconnected nodes does not depend as strongly on the scale-free nature of the degree distribution, with similar qualitative features found in both Fig. \ref{InitialValue.fig}C-D for the ER networks and Fig. \ref{BarabasiScaling.fig}C-D for the BA networks.  While the existence of hubs in the BA networks tends to give a higher probability of finding smaller values of $E_{ij}$ for disconnected nodes in comparison to ER networks, the most likely values of $E_{ij}$ are similar for disconnected nodes in either network topology (in contrast to the radically different distributions in for connected nodes).  We have considered only unweighted networks in this analysis, and allowing weighted edges  further complicates the analysis of the `typical' GEN between nodes unless a homogeneity assumption on the distribution of weights is likewise made.




\begin{figure}
\begin{center}
\includegraphics[width=.7\textwidth]{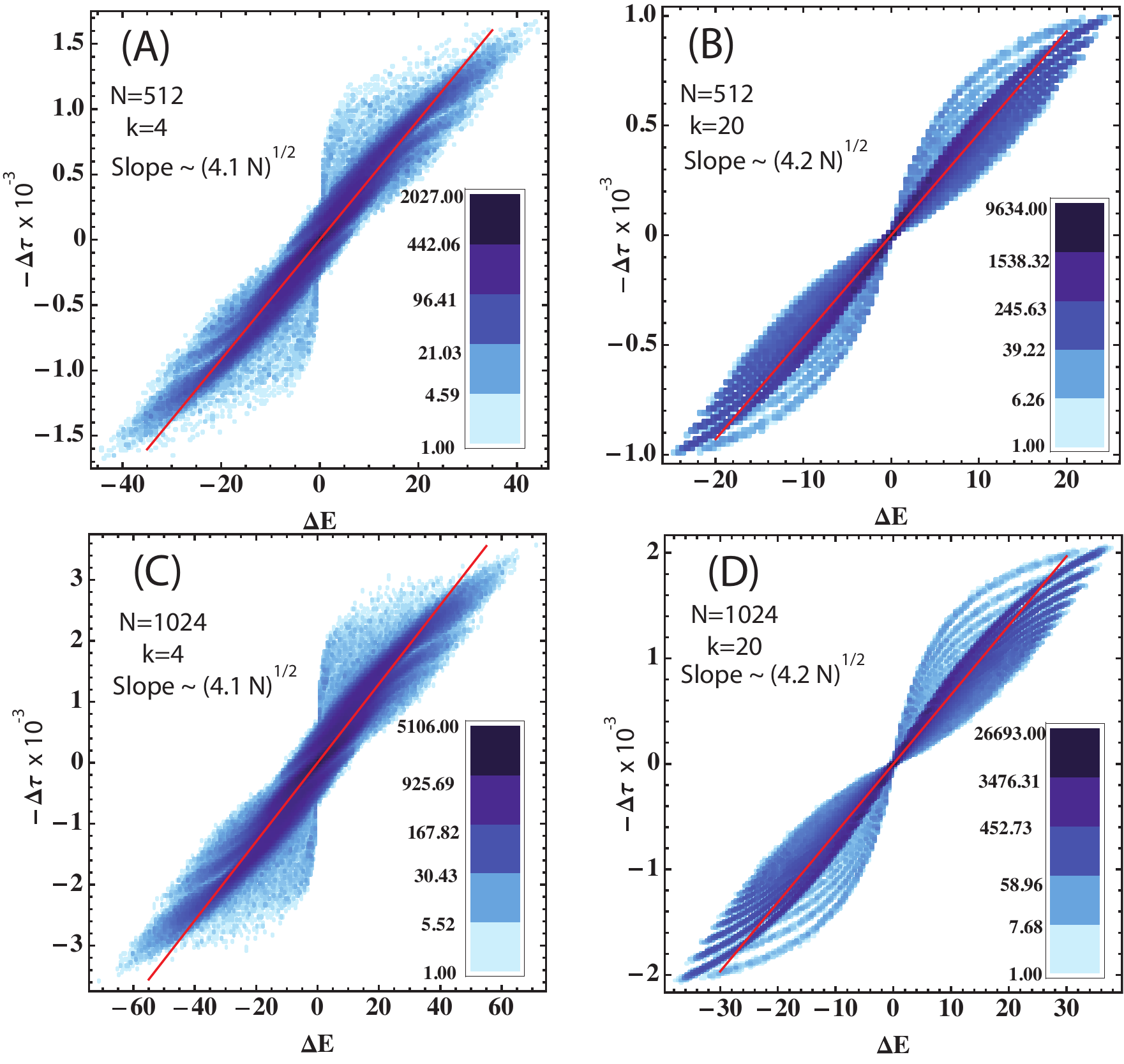}
\caption{Asymmetry in the Barab\'asi-Albert GENs $\delta E_{ij}=E_{ij}-E_{ji}$ compared with the asymmetry in the MFPTs for those networks, $\delta \tau_{ij}=\tau_{ij}-\tau_{ji}$.  In these density plots, darker colors correspond to a greater observed frequency of the same ($\delta E,\delta \tau$) pair.  Shown are two values of $N=512$ and 1024, as well as two values of $\la k\ra=4$ and 20.  It is visually clear that the asymmetry of the GENs are highly correlated to the asymmetry in the MFPT (with the slope of the best fit line indicated). \label{AsymmFig.fig}}  
\end{center}
\end{figure}

\section{Random Walks and the GENs}

\begin{figure*}
\begin{center}
\includegraphics[width=\textwidth]{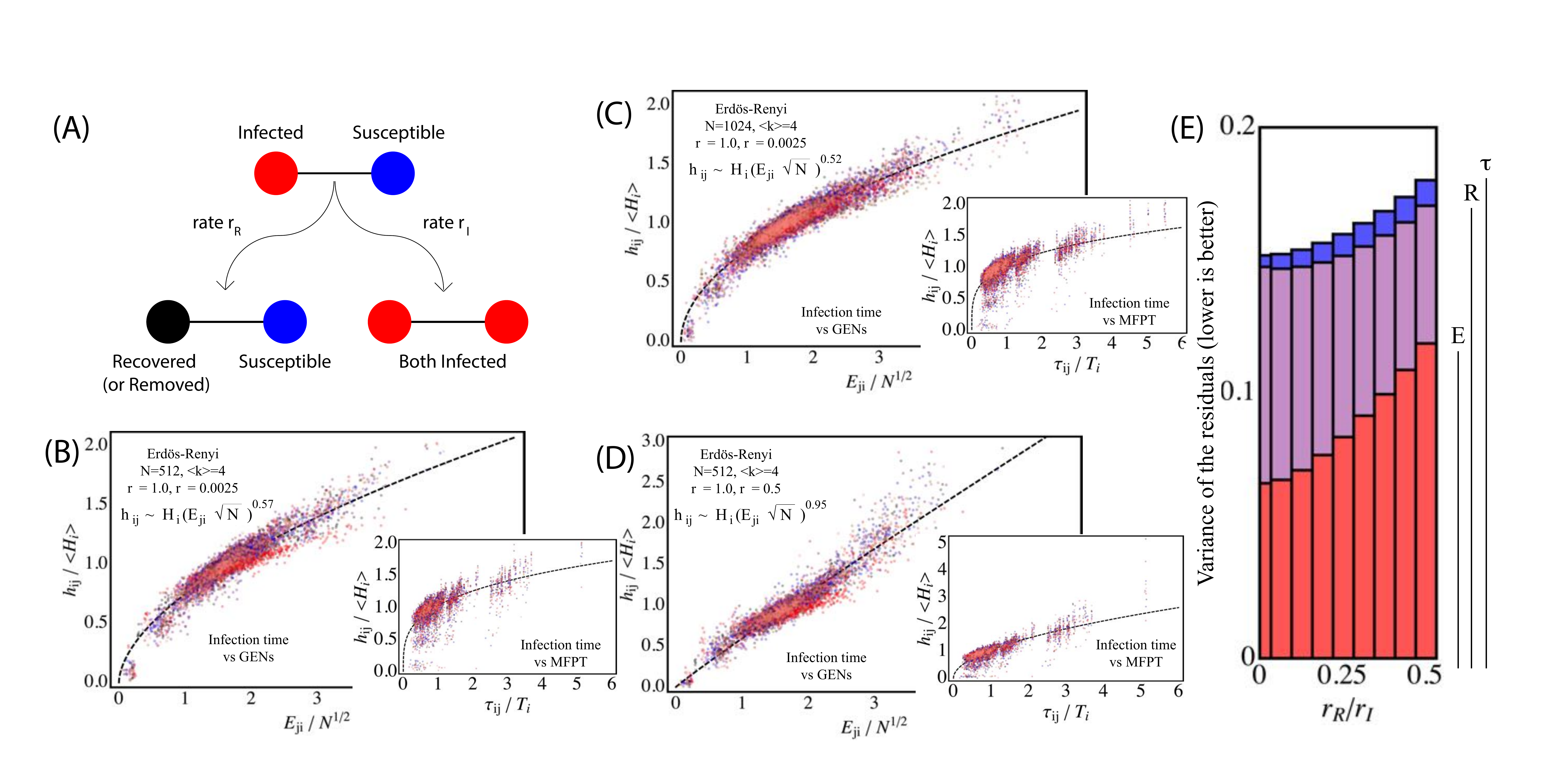}
\caption{The harmonic mean of the infection time of node $j$ with a single initially infected node $i$, $h_{ij}$ in an Erd\H{o}s-R\'enyi network. The SIR model is diagrammed in (A).  The main panels of (B-D) compared $h_{ij}$ with the GENs $E_{ji}$ while a comparison with the MFPT in a random walk $\tau_{ij}$ is shown in the inset, with different colors denoting a different initially infected node $i$.  In all, $h_{ij}$ is scaled by the average infection time of the nodes, $H_i$, and likewise the MFPT is scaled in terms of $T_{i}=(N-1)^{-1}\sum_{i\ne k}\tau_{ik}$.  (B) shows $N=512$ nodes with $r_R=0.0025R_I$. (C) $N=512$ and $r_R=0.5r_I$. (D) $N=1024$ and $r_R=0.0025r_I$, all with $\la k\ra=4$.   A functional relationship between $h_{ij}/H_i$ and $E_{ij}$ is immediately apparent in (B-D), which corresponds well to a simple power law for $r_R\ll r_I$, while the insets show that the MFPT is a poor predictor of average infection time.  The dashed lines show the best power law fit for the data, with the scaling indicated in the figure.  The functional behavior is not universal, depending on the size of the network and the SIR model parameters.  (E) shows the standard deviation of the residuals as a function of $r_R/r_I$ for a power law fit $h_{ij}/H_i\sim f(x_{ij})$ for $x_{ij}=E_{ij}$ (red), $R_{ij}$ (purple), and $\tau_{ij}/T_i$ (blue).  \label{SIRfig.fig}}
\end{center}
\end{figure*}

The mean first passage time (MFPT) from node $i$ to node $j$ ($\tau_{ij}$) are of interest in many contexts, and because of the limited resource represented by the random walker, it is worthwhile to see the relationship between the rate of travel between nodes and how `close' they are as measured by the GENs. {{Tetali}}\cite{TetaliJThProb91} has shown that the MFPT in an unbiased random walk can be computed directly from the resistance distance\cite{KleinJMathChem93,NewmanSocNet05} $R_{ij}$ with $\tau_{ij}=\frac{1}{2}\sum_l k_l(R_{ij}+R_{jl}-R_{il})$.  It is easily seen that the MFPTs are asymmetric ($\tau_{ij}\ne \tau_{ji}$), in general, as it is easier to reach a high-degree node than a low degree node (much like the asymmetry in the GENs with $E_{ij}\ne E_{ji}$).  We intuitively expect that if node $j$ feels `close' to node $i$ (small $E_{ij}$) but node $j$ does not (large $E_{ji}$), a random walker will pass more readily from $j$ to $i$ than from $i$ to $j$.   We therefore compare $\delta E_{ij}=E_{ij}-E_{ji}$ to the difference in the MFPT between nodes $\delta \tau_{ij}=\tau_{ij}-\tau_{ji}$ in Fig. \ref{AsymmFig.fig} and empirically find the asymmetry in the MFPT is highly correlated with the asymmetry in the GENs, with an empirical scaling of $\delta \tau_{ij}\sim -\delta E_{ji} \sqrt{\alpha N}$ with $N$ the number of nodes in the network and $\alpha\sim 4$ a topology-dependent constant.  
The fact that $\delta \tau_{ij}\propto \delta E_{ji}$, even when there are no direct connections between $i$ and $j$, indicates that the GENs are able to capture the importance of the global network topology even for distant nodes.  It is interesting to note that while the relative proportionality appears to be strongly dependent on the size of the network and only weakly dependent on the average degree, it is visually apparent that $\la k \ra$ sets the scale of the fluctuation statistics from the best fit line.  For $\la k\ra=4$, the asymmetry density is relatively disperse (Fig. \ref{AsymmFig.fig}(A) and (C)) with no obvious structure, while for $\la k\ra=20$ there is clear clustering of the density about curved bands (Fig. \ref{AsymmFig.fig}(B) and (D)).  

\begin{figure}[tbp]
\begin{center}
\includegraphics[width=.7\textwidth]{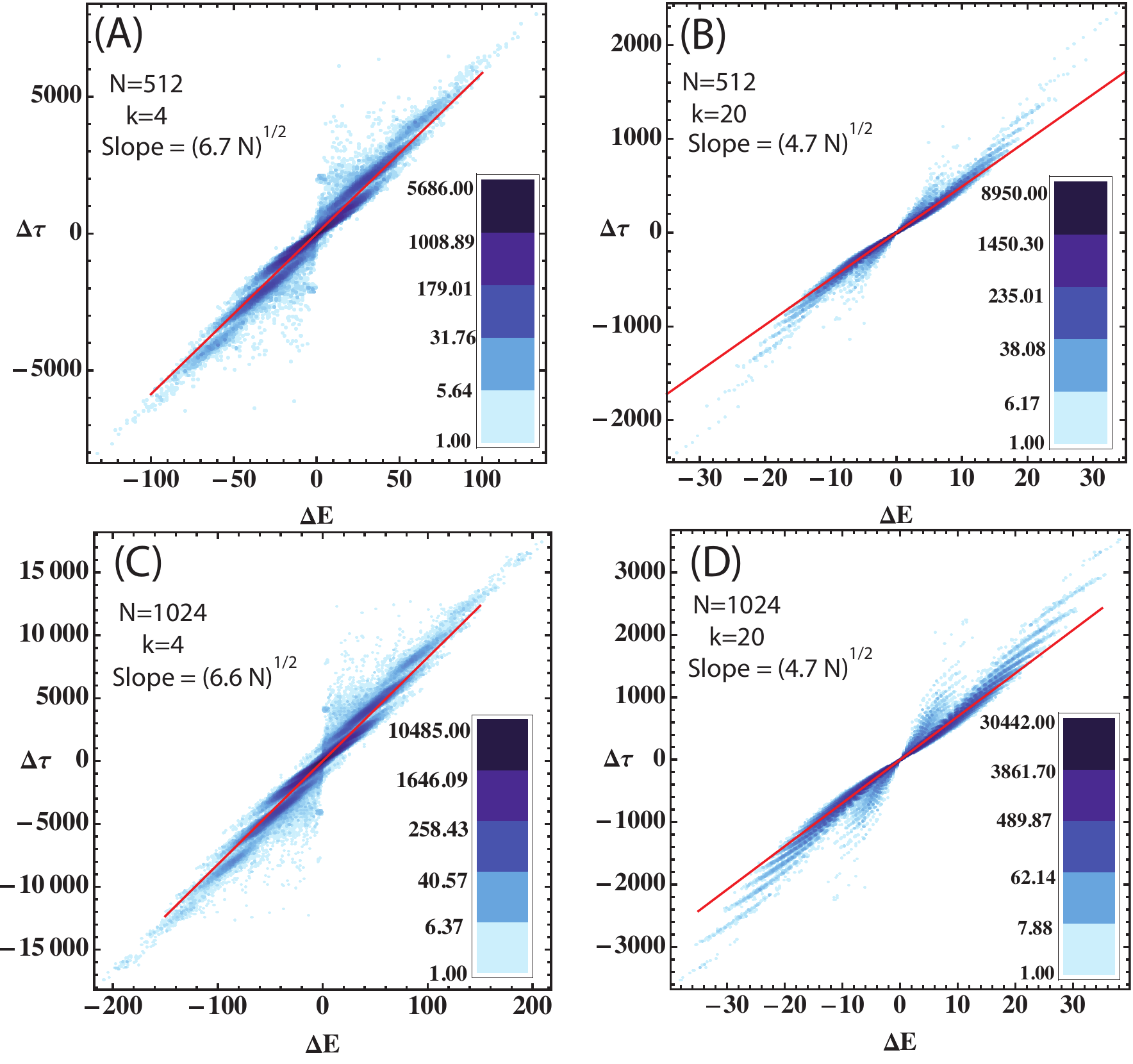}
\caption{Asymmetry in the Erd\"os-Renyi GENs $\Delta E_{ij}=E_{ij}-E_{ji}$ compared with the asymmetry in the MFPTs for those networks, $\Delta \tau_{ij}=\tau_{ji}-\tau_{ij}$.  The figure labels are identical to those in Fig. \ref{AsymmFig.fig}, with a similar behavior in scaling and variation from the best fit line.  
}
\label{AsymmFigAppendix.fig}
\end{center}
\end{figure}

A similar result, using the same values of $N$ and $\la k\ra$ for a Erd\H{o}s-R\'enyi network, is shown in Fig. \ref{AsymmFigAppendix.fig} and displays the same qualitative features:  the asymmetry in the GENs are highly correlated with the asymmetry in the mean first passage times, and there is an apparent scaling of $\delta E_{ij}\sim \delta \tau_{ij}/\sqrt{\alpha_{ER} N}$ with $\alpha\sim 4-7$ depending on the average degree.  This coefficient is on the order of the scaling coefficient for the BA networks in Fig. \ref{AsymmFig.fig}, with $\alpha_{BA}\sim 4$.  While the structure of the deviation from the best fit line is interesting, note that the logarithmic scale of the coloring means that these extreme outliers are still relatively rare compared to the much higher density near the best fit line.  The overall correlation between the asymmetries of the GENs and the MFPTs indicates that $\delta E$ is a meaningful quantity, capturing the topological details of the network.  We have found that our iterative method for computing the GENs (implemented in C++) converges far more rapidly than directly computing the MFPT via a matrix pseudo-inversion\cite{KleinJMathChem93} for the networks we have considered (using the `pinv' function in Matlab). 

\section{An application to epidemic spreading}

The spreading of an epidemic has been studied by many authors and in a wide range of contexts\cite{AndersonBook,BallMathBio2008,LegarioPRE11,HethcoteSIAMRev00,SatorrasPRL01}, with the SIR model being one of the simplest and most commonly used.  The susceptible-infected-recovered (SIR) model assumes that a population of susceptible individuals becomes infected due to interactions with previously infected individuals, and infected individuals may recover and become non-infectious.  A simple schematic of the SIR model is shown in Fig. \ref{SIRfig.fig}(A), with infections occurring at a constant rate rate, $r_I$, when individuals interact and the recovery at constant rate, $r_R$.  A number of more complex models have been considered extensively for a homogeneously mixed population of individuals \cite{HethcoteSIAMRev00}, but non-uniform interactions between individuals, represented by networks, can have a profound impact on the dynamics of epidemic spreading in the SIR model\cite{NewmanPRE02,SatorrasPRL01,BallMathBio2008}.  The existence of epidemic thresholds\cite{NewmanPRE02,MillerJMB11} for homogeneous networks (or the lack thereof for scale free networks\cite{SatorrasPRL01}) are well-studied global quantities of interest, while more local quantities such as the probability of a particular node $i$ becoming infected, sparking an epidemic\cite{MeyersBAMs07}, and quarantine or immunization strategies\cite{LegarioPRE11,SatorrasPRE02} have also been examined.

While it is clearly useful to understand the global properties of the epidemic (such as the expected number of infected individuals), a particular individual $j$ may also be interested in its own probability of becoming infected given the current state of the disease  
and may reasonably be less concerned if no neighbors are infected than if many neighbors are infected.  However, it is not straightforward to analytically calculate how long the disease will take to reach $j$ from any point in the network, and it would be useful to have a measure for how `close' the epidemic is from an individual node.  
If the infection begins with a single node $i$, we expect that the disease will more rapidly propagate to nodes for which $i$ feels topologically close, and it is therefore worthwhile to compare the infection times in a SIR epidemic with the GENs as a proxy for closeness.  To see the relationship between infection time and closeness, we simulate an SIR epidemic, using Gillespie dynamics\cite{GillespieJCompPhys76} on an Erd\H{o}s-R\'enyi graph (with a uniform probability of connection and each node having $\la k\ra=4$ average degree). This allows us to determine the harmonic mean of the infection time of a node $j$ given an initial infection at $i$ over the $K$ simulations, $h_{ij}^{-1}=\sum_{k=1}^K [t^{(k)}_{i\to j}]^{-1}$ (the harmonic mean is chosen to avoid diverging infection times in simulation $k$, $t_{i\to j}^{(k)}$, where the disease dies out before $j$ is infected).  Because the GENs do not naturally include the timescales of the system dynamics ($r_I$ and $r_R$) we normalize the harmonic mean by $H_{i}=\frac{1}{N-1}\sum_{j\ne i} h_{ij}$, which is the `typical' time at which a node becomes infected if the disease originates with $i$.  

In Fig. \ref{SIRfig.fig}(B-D) we compare $h_{ij}/H_i$ to $E_{ji}$ (how close node $i$ feels towards node $j$) for a network of 512 nodes and $r_R=0.0025r_I$ in (B); $N=1024$ and $r_R=0.0025r_I$ in (C); and $N=512$ and $r_R=0.5r_I$ in (D).  In all cases, $h_{ij}/H_i$ tends to increase with $E_{ji}$ (as is expected:  nodes $i$ feels close to are infected faster), although the functional dependence of the infection time on the topological closeness depends on the parameters in a non-universal way.  We also show in the insets that the scaled mean first passage time between nodes in a random walk ($\tau_{ij}/T_{i}$) does a surprisingly poor job of predicting the infection time in comparison with the GENs.  This is likely due to the significant differences between the SIR model and a random walk:  if a walker departs from node $i$ at timestep $t$, he cannot depart from $i$ again at time $t+1$.  In effect, there is only one `infectious' node at a time  in a random walk which makes the MFPT a poor estimator of the infection time in a SIR simulation.   The resistance distance between nodes $R_{ij}$ provides a fit quality similar to that of the scaled MFPT (data not shown).  In Fig. \ref{SIRfig.fig}(E), the quality of the fit is quantified using the standard deviation of the residuals of the fit.  In all cases, the GENs are much more highly correlated with the average time of infection between pairs of nodes in the network.  Fig. \ref{SIRfig.fig} indicates that the GENs $E_{ij}$ can more meaningfully capture the impact of network topology on the dynamics of epidemic spreading than other global measures of pairwise `closeness' between nodes.

%
%
%

\section{Measuring a personalized importance}
\label{ImportanceSection}

The GENs incorporate a simple idea of what is meant by the `closeness' between nodes in a network where limited resources are shared, and we expect that a node $j$ that feels `close' to node $i$ (having small $E_{ij}$) considers node $i$ to be `important' in some sense.  We may therefore regard the inverse of the closeness between nodes ($\psi_{ij}=E_{ij}^{-1}$) as an un-normalized personalized measure of importance, allowing a ranking of all nodes in the network {{from the perspective of the node $j$}}.  Because $\psi_{ij}$ measures the importance of $i$ from a particular node $j$ (rather than the network at large), it is not equivalent to a centrality measure.

\begin{figure}[htbp]
\begin{center}
\includegraphics[width=.3\textwidth]{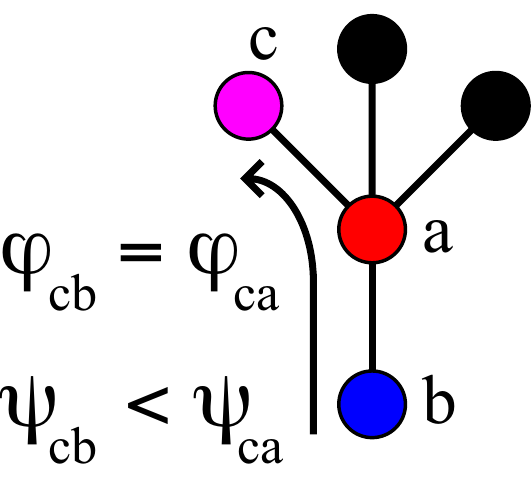}
\caption{In this simple network, $\phi_{cb}=\phi_{ca}$, meaning the importance node $b$ assigns to $c$ (with no direct connection between them) is identical to the importance $a$ assigns to $c$ (which share a direct connection).   Intuitively, we would expect the lack of direct connection to imply $b$ feels $c$ is less important (due to the lack of direct connection $w_{bc}=0$).  This condition is satisfied using the nonlinear importance, with $\psi_{cb}<\psi_{ca}$.}  
\label{LinearFailure.fig}
\end{center}
\end{figure}

To gain insight into the meaning of the personalized importance, Eq. \ref{GENdef} can be expanded in the limit of large $E_{ij}$ for $i\ne j$ (a valid expansion for $k_i=k$ shown in the SI, and is observed to be reasonable for more complex networks), yielding $E_{ij}^{-1}\approx W_j^{-1}[w_{ij}^2+\sum_{i\ne l\in C_j'}w_{ij}E_{ij}^{-1}]+O(E_{ij}^{-2})$.  We can use this lowest-order expansion to define the linearized importance assigned by node $j$ towards node $i$ ($\phi_{ij}$) as 
\begin{eqnarray}
\j_{ij}\equiv\frac{w_{ij}^2}{W_j}+\frac{1}{W_j}\sum_{{i\ne l\in C_j}}{w_{jl}}{}\j_{il},\label{importanceEq}
\end{eqnarray}
where $\j_{ii}$ is undefined (since $E_{ii}=0$).  Eq. \ref{importanceEq} provides a natural interpretation of the meaning of personalized importance.  The importance $j$ assigns to $i$ is a combination of two terms:  a weighted average of the importance his neighbors assign to $i$, and an importance of $w_{ij}$ assigned in the case of a direct connection between $i$ and $j$.  
Defining $({\bm{\phi}})_{ij}=\phi_{ij}$ and $({\mathbf{L}})_{ij}=W_j\delta_{ij}-w_{ij}$ the graph Laplacian\cite{BabicIntJQantChem02,LaplacianMachineLearning,KleinJMathChem93}, Eq. \ref{importanceEq} can be written $({\bm{\phi}}{\mathbf{L}})_{ij}=w_{ij}^2-w_{ij}\phi_{ij}$, relating the linearized importance directly to the graph Laplacian.  
%
%
Despite the simplicity of this interpretation of the linearized importance, $\phi_{ij}$ is of limited use in unweighted networks (diagrammed in Fig. \ref{LinearFailure.fig}: if $w_{ij}= \{0,1\}$, Eq. \ref{importanceEq} reduces to $k_{j}\times \phi_{ij}=\sum_{l\in C_j}\phi_{il}$ with the solution $\phi_{ij}=1$ for all connected nodes.  
Only in the context of weighted networks is a non-trivial measure of importance obtained.  The full nonlinear expression for the GENs (which incorporates higher order terms than the linear approximation) does not suffer from this difficulty and can still determine meaningful measures of importance between nodes.  

\begin{figure}
\begin{center}
\includegraphics[width=0.6\textwidth]{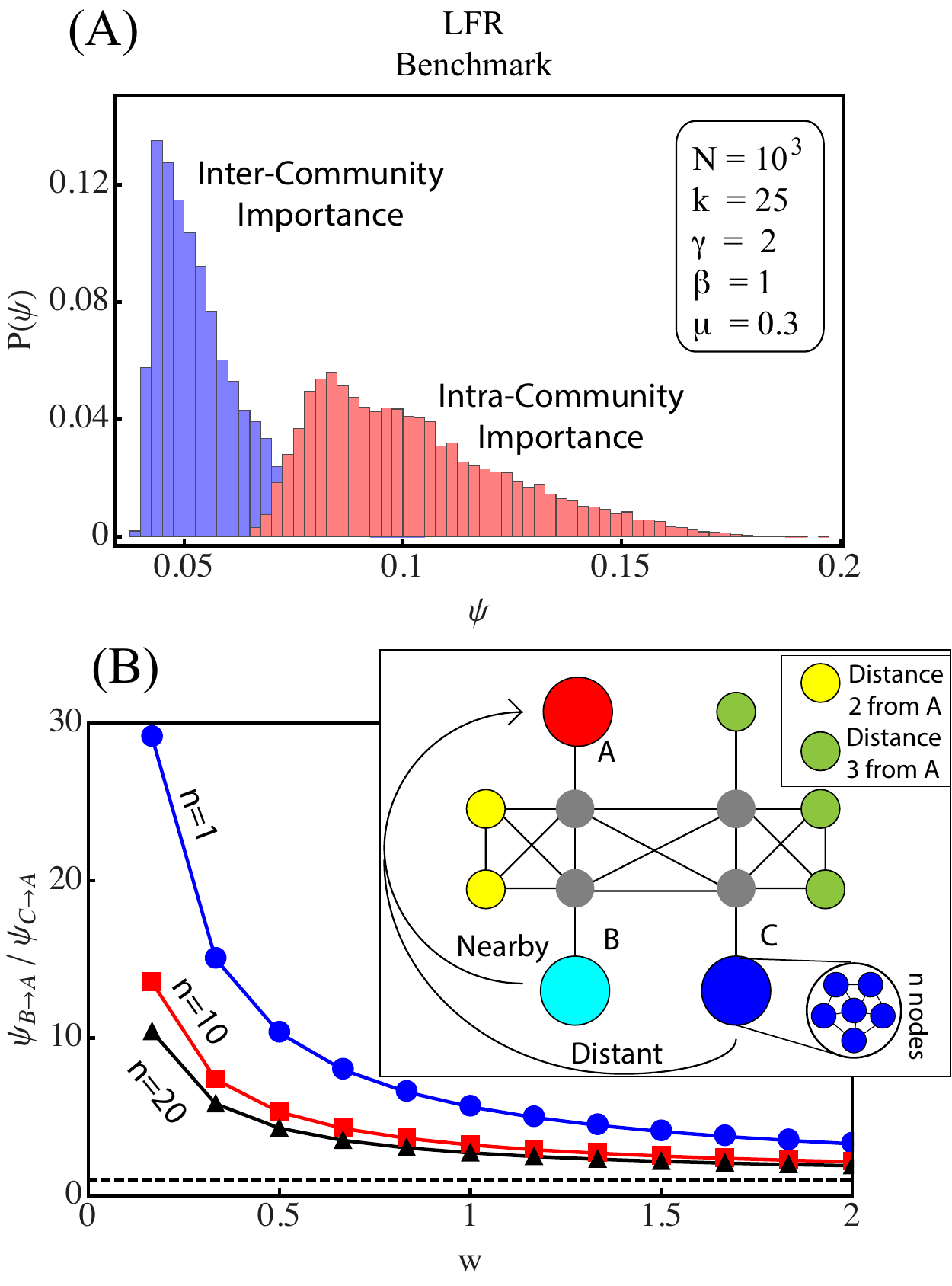}
\caption{(A)  The GENs applied to the LFR benchmark\cite{LFR}.  The red shows the distribution of importance for nodes $i$ and $j$ that are in the same community but do not share a direct connection. The blue shows the distribution for those in different communities (and still sharing no direct connection).  Due to the high density of links inside of the communities, the GENs accurately indicate that $\psi_{ij}$ is likely to be larger if $i$ and $j$ are in the same community.  (B) In the inset,  we diagram a simple network topology where cliques are in greater proximity to one another but still have no neighbors in common.  Each circle represents a clique of $n$ nodes (fully connected with each edge having unit strength) and each edge an all-to-all connection between cliques with each edge having weight $w$.  Clique B is in close proximity to auxiliary nodes with integer distance 2 from A (highlighted in yellow) while clique C is in closer proximity to auxiliary nodes of integer distance 3 (highlighted in green).  In the main figure, $\psi_{ij}$ indicates that A is more important from the perspective of B than from C for varying values of $n$ and $w$ due to the closer proximity of B's neighbors to A than C's neighbors, with $\psi_{AB}/\psi_{AC}>1$.\label{personalImportance.fig}}
\end{center}
\end{figure}

 The usefulness of the nonlinear importance $\psi_{ij}=E_{ij}^{-1}$ on a network can rapidly determine meaningful relationships between nodes in complex networks.  To illustrate this, we consider the benchmark of Liancichinetti, Fortunato, and Radicchi (LFR)\cite{LFR}, which constructs a network of communities of variable sizes $n$ (distributed as $P(n)\propto n^{-\beta}$), a scale-free distribution of the nodes (with $P(k)\propto k^{-\gamma}$), and which is characterized by the mixing parameter, $\mu$, as the fraction of inter-community edges.  We have previously shown\cite{MorrisonPLoS12} that the GENs can be used to detect the community structure underlying this benchmark.  When measuring the importance of a node, a global measure of centrality will generally focus on nodes with high degree, but due to the heterogeneous density of edges between communities, we expect a meaningful definition of the importance $j$ assigns to $i$ to differ significantly depending on if $i$ and $j$ are in the same community.  Note that the determination of the GENs does not require {\em{a priori}} knowledge of the community structure.  In Fig. \ref{personalImportance.fig}(A), we determine the distribution of importance $\psi_{ij}$ between nodes $i$ and $j$ that do not share a direct connection ($w_{ij}=0$) for nodes within $i$'s community (red) and outside of $i$'s community (blue).  There is an immediately apparent difference in the distributions, with a greater probability of a high importance if $i$ and $j$ are in the same community due to the increased number of shared neighbors (even in the absence of a direct connection).  Increasing the LFR parameter $\mu$ (which increases the number of edges between communities) reduces the difference in the distributions, but varying the other system parameters has only a minor impact on the clear distinction between the two distributions (data not shown).  A large overlap between neighbors is not required to accurately detect meaningful differences in the importance between pairs of nodes, though: in Fig. \ref{personalImportance.fig}(B) we show cliques of $n$ nodes with $A$, $B$, and $C$ having no neighbors in common, but due to the network topology, $A$ is closer to $B$ than to $C$ (depicted in the inset of Fig. \ref{personalImportance.fig}(B)).  The GENs determine that $B$ feels $A$ is more important than $C$ does.  $\psi_{ij}$ incorporates direct connections between nodes, proximity in the network, and shared neighbors into a meaningful measure of personalized importance.

\section{Global importance and Erd\H{o}s centrality}
\label{CentralitySec}

 \begin{table}
 \begin{tabular}{@{\vrule height 10.pt depth3pt  width0pt}lrcccc}
 $k_i$ & {\small{Degree\cite{OpsahlSocNet10} }} & {\small{Node Degree}}\\
 \hline
$Pr_i$ &{\small{PageRank\cite{FranceschetComACM11} }} &  {\small{$i^{th}$ component of first eigenvector of $\Bv$}}\\
\hline
 $b_i$ & {\small{Betweenness\cite{NewmanSocNet05}}}  & {\small{ times $i$ is visited in a random walk}}\\
& {\small{Centrality}} & {\small{$\propto \sum_{j\in C_i}\sum_{l,m\ne i} |R_{il}+R_{jm}-R_{im}-R_{jl}|$.}}\\
\hline
$c_i$ &{\small{Random Walk \cite{NohPRL04} }}& {\small{The rate information flows from other}} \\
&{\small{Centrality}}  & {\small{nodes to $i$:  $c_i^{-1}-c_j^{-1}=\tau_{ji}-\tau_{ij}$}}\cr
\hline
 \end{tabular}
  \caption{
 Some common methods  of measuring the importance of a node $i$.   These are compared directly to the Erd\H{o}s Centrality $\Psi_{i}=\sum_{l\in C_{i}} E^{-1}_{il}$ in Fig. 
 4 of the main text.  This list is not exhaustive, and other measures of centrality are possible\cite{OpsahlSocNet10,BorgattiSocNet05}.  $R_{ij}$ denotes the resistance distance\cite{KleinJMathChem93,NewmanSocNet05} between nodes $i$ and $j$,  $\tau_{ij}$ the mean first passage time between $i$ and $j$ in a random walk, and $\Bv_{ij}=(1-d) w_{ij}/W_j+d/N $ is the matrix of transition probabilities for PageRank (where $d$ is the teleport probability of visiting a non-neighbor in PageRank).  Throughout the paper, each measure of centrality is normalized such that the sum of all centralities in the network is 1.
 }
 \label{centrality.tab}
 \end{table}

\begin{figure}
\begin{center}
\includegraphics[width=\textwidth]{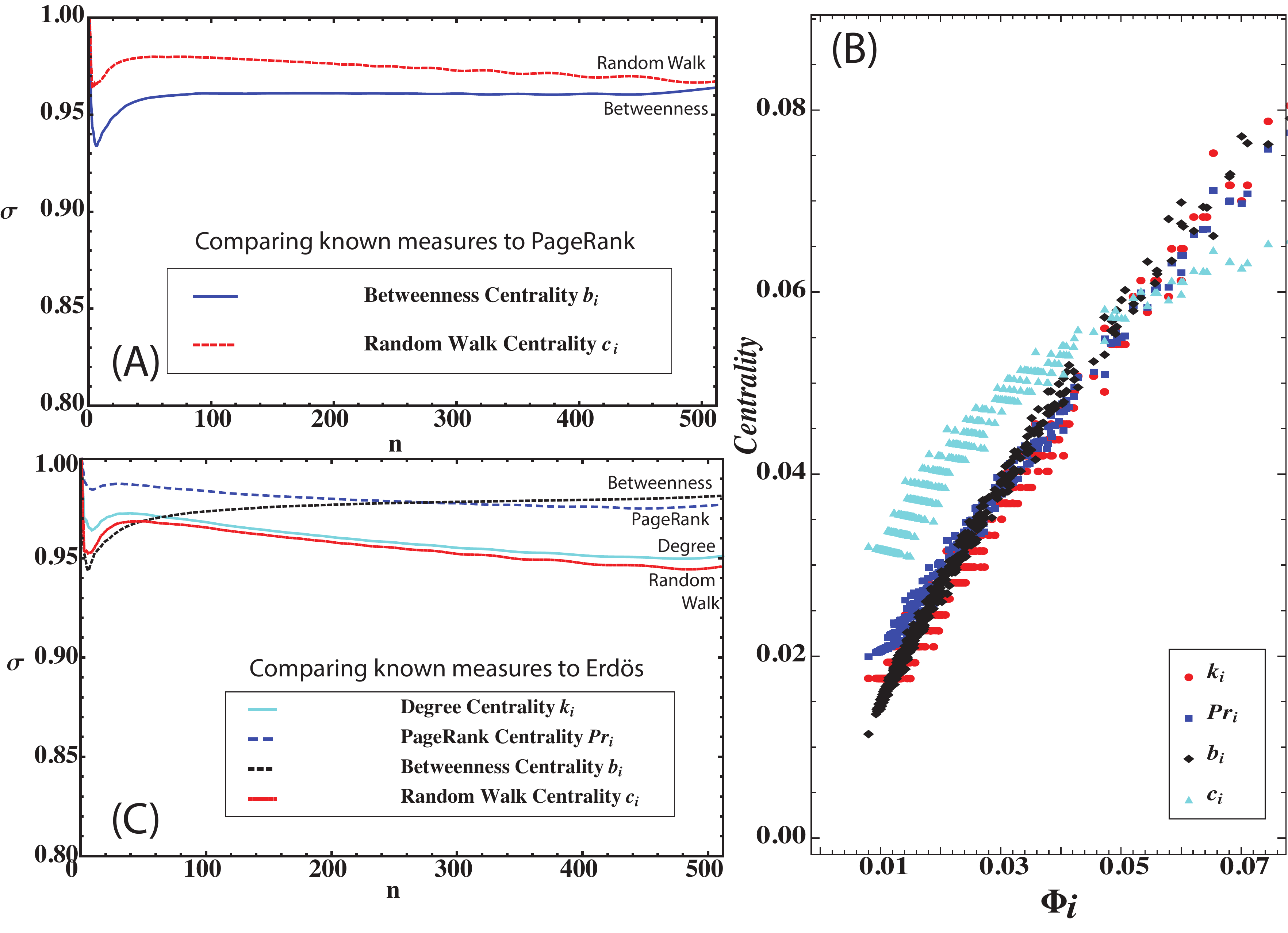}
\caption{Centrality for a Barab\'asi-Albert network with $\la k\ra=20$.  (A) uses the intersection metric $\sigma_{XY}(n)$ to estimate the similarity in top-$n$ lists between PageRank and two centrality measures listed in Table 
1 in the Supplementary information.  While there is good correspondence between these measures, each measure gives a slightly different ordering for small $n$.  (B)  The Erd\H{o}s centrality (x-axis) compared to the 4 common centrality measures 
(y-axis) shows an obvious positive correlation overall.  Circles shows degree centrality, squares PageRank, diamonds betweenness centrality and triangles random walk centrality.  The clustering of some data near discrete values is due to the heterogeneity of the Erd\H{o}s centrality for nodes of equal degree (this effect is more pronounced in Fig. \ref{CentralityFigAppendix.fig} with $\la k\ra=4$).  (C)  shows the intersection metric between the top $n$ elements of the Erd\H{o}s centrality ($\mathbf{o}_E(n)$) and the top $n$ elements of the other centrality measures for varying $n$.  The GENs are about as consistent with known measures of centrality as these measures are amongst themselves (shown in (A)).  \label{ErdosCentrality.fig}}
\end{center}
\end{figure}

Having defined a pairwise measure of the importance a node $j$ assigns to $i$ using $\psi_{ij}$ (or the linearized $\phi_{ij}$), we naturally expect that we can leverage this definition into a global measure of the importance of node $i$.  There already exist a wide variety of methods for measuring centrality from a global perspective, including degree\cite{OpsahlSocNet10,NewmanSocNet05,BorgattiSocNet05}, PageRank\cite{BorgattiSocNet05,FranceschetComACM11}, random walk\cite{NohPRL04}, and betweenness centrality\cite{NohPRL04,NewmanSocNet05} (briefly described in Table \ref{centrality.tab}).
  Each measure ranks high-degree nodes above low-degree nodes but take the global network topology into account in different ways.   These methods produce qualitatively similar but quantitively different node rankings, as reflected by the fractional intersection between the top-$n$ orderings\cite{RankingSimilarity}, $\sigma_{XY}(n)=\frac{1}{n}\sum_k |\mathbf{o}_X(k)\cap \mathbf{o}_Y(k)|/k$, with $\mathbf{o}_X(k)$ the top-$k$ ordering using method $X$.  $\sigma_{XY}(n)$ is shown comparing the ordering due to PageRank with that found using the betweenness and random walk centrality measures for a Barab\'asi-Albert network topology in Fig. \ref{ErdosCentrality.fig}(A).  It is clear that there is an immediate drop to $\sigma \gtrsim 0.95$ for small $n$ for these well known centralities (i.e. good but not perfect agreement on the top few nodes), after which the methods tend to vary slowly or not at all for larger top-$n$ lists.

To convert our personalized importance measures into a single global measure, we define $\Psi_i=\sum_{l\in C_i}\psi_{il}$ as the sum of the importance the neighbors of $i$ assign to it (akin to the approach of Ref. \cite{LeichtPRE06}).  For an unweighted network (with $\phi_{ij}=1$ for all $i$ and $j$), the linearized $\Phi_i=\sum_{j\in C_i}\phi_{ij}=k_i$ is equivalent to the degree centrality.  In Fig. \ref{ErdosCentrality.fig}(B), we compare $\Psi_{i}$ to the other measures of centrality in Table 1 for a single realization of a Barab\'asi Albert network with $N=512$ and $\la k\ra=20$.  In all cases, there is an obvious correlation between these measures of centrality but with significant difference between centrality measures in some cases for both central and non-central nodes alike.  Degree and random walk centrality both tend to be strongly clustered around desecrate values, leading to the banded structures seen in Fig. \ref{ErdosCentrality.fig}(B), and the difference between these and PageRank seen in Fig. \ref{ErdosCentrality.fig}(A).  The Erd\H{o}s centrality takes the global topology into account differently than these measures of centrality (similar to the betweenness and PageRank centrality).  When comparing the Erd\H{o}s centrality to those in Table. 1, the similarity between the orderings from most- to least-important\cite{TopicPageRank,RankingStability}, using different measures is of particular interest\cite{GhoshalNatCom11,BlummPRL12}.   We compare the Erd\H{o}s centrality ordering to the other measures of centrality using $\sigma_{XY}(n)$\cite{RankingSimilarity} in Fig. \ref{ErdosCentrality.fig}(C) and see that the Erd\H{o}s centrality is consistent with other measures of centrality:  a sharp drop initially with slow variation for larger $n$ and $\sigma(n)\gtrsim 0.95$ throughout.  Despite the different formulations between the Erd\H{o}s centrality and PageRank, the top-$n$ list for $\Psi_i$ compares best to the list from $Pr_i$ (dashed purple line) for high degree nodes but begins to agree better with betweenness centrality when less central nodes are also included.

\begin{figure}[tbp]
\begin{center}
\includegraphics[width=\textwidth]{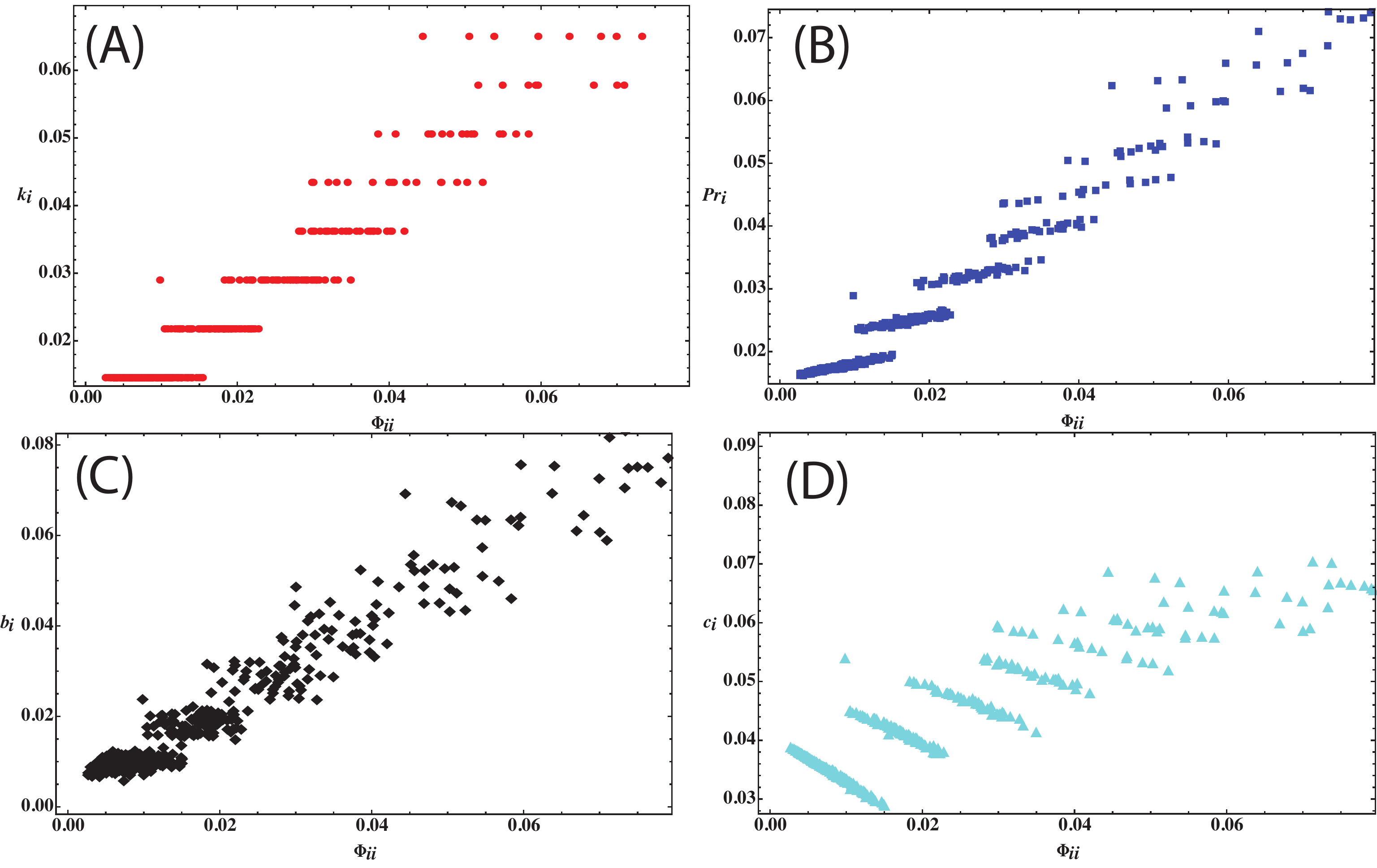}
\caption{Comparing the Erd\H{o}s centrality with other measures of centrality for Barab\'asi-Albert networks with a lower average degree of $\la k\ra=4$.  The symbol color and shape are the same as in Fig. \ref{ErdosCentrality.fig}.  There is pronounced clustering in the data due to the strong dependence the degree and PageRank measures of centrality have on the node degree, and the Erd\H{o}s centrality will sometimes label a node as more important than another node with a somewhat higher degree.  The overall rankings are still consistent, with high degree nodes being typically more important.   
}
\label{CentralityFigAppendix.fig}
\end{center}
\end{figure}

We have also computed the Erd\H{o}s centrality and other measures of centrality for a BA network with $N=512$ and $\la k\ra=4$.  
While the discreteness of the degree distribution is still apparent in Fig. \ref{ErdosCentrality.fig}B (with the tight clustering of points about discrete values), it is has a greater impact in Fig. \ref{CentralityFigAppendix.fig}, where $k$ attains fewer values due to the significantly smaller average degree.  The general trends are qualitatively similar to that in Fig. \ref{ErdosCentrality.fig}, though:  the betweenness centrality (Fig. \ref{CentralityFigAppendix.fig}(C)) depends on the degree in a qualitatively different manner than any other measure (and is more clearly correlated to the Erd\H{o}s centrality for low-ranked nodes).  The Erd\H{o}s centrality does not appear to be linearly related to the random walk centrality (although they are still correlated), and clear bands are observed due to nodes of equal degree for both the degree and PageRank measures of centrality.

It is interesting to note that $\psi_{ij}$ can define a modified connectivity matrix, incorporating both direct connections in the adjacency matrix and non-local paths between nodes.  A random walk performed with a transition probability $\Bv'_{ji}=\psi_{ij}/\sum_{l\ne i}\psi_{lj}$ (with the convention $\Bv_{ii}'=0$, meaning the walker never remains at $i$) has a similar interpretation:  a walker at node $j$ has a relatively high probability of moving to node $i$ if they are directly connected in the original network (if $w_{ij}>0$), but has a non-zero probability of jumping to a disconnected node.  The matrix $\Bv'$ can be compared to the PageRank transition probability matrix $\Bv$, which has a uniform probability of teleporting to any node in the network (regardless of the network topology).  $\Bv'$ has a jump probability that is not purely random but rather preferentially lands on nodes that $j$ finds important, in contrast to PageRank's uniform teleport probability.  An alternate Erd\H{o}s measure of centrality can be defined as the largest eigenvector of the matrix $\Bv'$ (akin to PageRank, the steady state probability of being found at a node $i$ under the transition probability $\Bv$).  However, the computational efficiency of the simple sum $\Psi_{i}=\sum_{j\in C_i} \psi_{ij}$ and its clear correlation with known measures of centrality (seen in Figs. \ref{ErdosCentrality.fig} and \ref{CentralityFigAppendix.fig}) makes our current definition of the Erd\H{o}s centrality preferable.

Because the GENs can be used to describe random walks efficiently while prescribing a topology-dependent teleportation via the matrix $\Bv'$ (with $\Bv'_{ij}=\psi_{ij}/\sum_{l\ne i}\psi_{lj}$ discussed in Sec. \ref{CentralitySec}), they may be of use in future studies as well.  Walkers on $\Bv'$ will tend to remain trapped in regions of locally dense edges, and implementing such a walk in community detection methods may increase the accuracy of the determined partition\cite{PonsCIS05}.  Similarly, the modified walk may be of use in modeling systems where jumps to non-neighboring (but topologically close) nodes is desirable, such as epidemic spreading in a social network.  Epidemics can be been modeled\cite{BallMathBio2008} as spreading with some probability to direct contacts (due to the assumed high probability they interact directly) and with a lower probability `jumping' to nodes without a direct contact (with non-neighbor transmission possible due to rare transient interactions).  We could expect that non-neighbor interactions will not have a uniform probability, but rather that each node is more likely to transiently interact with friends-of-friends (or individuals s/he feels topologically close to).  Epidemic spreading on $\Bv'$ will capture our expectation that transmission to friends with whom we have regular contact is most likely, but transmission to friends of friends is more likely than transmission to very distant people in the social network.

\begin{figure}
\begin{center}
\includegraphics[width=\textwidth]{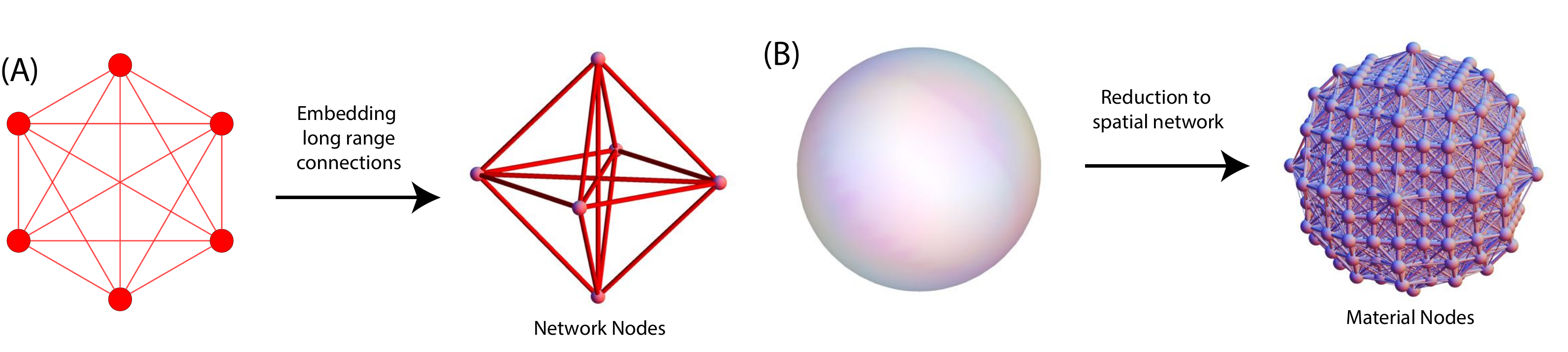}
\caption{Schematic diagram of the continuum model with long range connections.  (A) shows the reduction of the continuum to a network of material nodes, with the weight between them dependent only on their distance, $w(\xv,\xv')=e^{-\lambda |\xv-\xv'|/a}$.  (B) shows the embedding of long range connections between network nodes, with weight specified by the network rather than depending on the distance between the nodes. \label{SpatialSchematic.fig}}
\end{center}
\end{figure}

\section{GENs in the continuum limit}

We showed in the previous sections that the Generalized Erd\H{o}s Numbers can be useful in describing complex networks a variety of contexts by incorporating the global network topology into a measure of closeness between nodes.  Many physical networks exist in a metric space\cite{BarthelemyPhysRep2011}, such as wireless sensor networks\cite{WirelessSensor,GeographicNetwork} or transportation\cite{KurantPRL05,MontisEnvPlan07} networks, where the strength of the direct interaction between two nodes depends on their locations.  In many contexts, such as transportation networks\cite{BalcanPNAS09}, distance-independent connection strengths may coexist with geometrically defined links (for example, surface mobility versus air travel).  It is natural to wonder what impact this may have on the GENs, and how topological closeness is influenced by physical proximity.  We assume that a network of long-range connections has been externally defined, and that each node in that network can be assigned to a location in some $d$ dimensional space (a schematic is shown in Fig. \ref{SpatialSchematic.fig}A).  Each node $i$ in the original, distance-independent network (of $N$ nodes) occupies a location $\yv_i$, with an interaction strength $w_{ij}$.  We refer to these nodes as `network nodes', each of which has at least one direct connection due to the connectivity of the original network (referred to as a long-range connection).  To introduce geometry into the problem, we also define a network of $N^*$ `material nodes' into the network at locations $\{\xv_j\}$, whose interactions are {{purely}} geometric (schematically diagrammed in Fig. \ref{SpatialSchematic.fig}B for a spherical geometry).  The distance-dependent interaction between the nodes of any type (both network and material nodes) at locations $\zv$ and $\zv'$ is given by $u(\zv,\zv')$, with the constraint that $u(\zv,\zv')=u(\zv',\zv)$ to ensure the network remains undirected. Note that distance-dependent connections are drawn between network nodes as well, so the total weight between $\yv_i$ and $\yv_j$ is $w_{ij}+u(\yv_i,\yv_j)$.  If $N^*=0$ or $u(\xv,\xv')=0$ the physical geometry is irrelevant and the original network is recovered, while if $N=0$ or $w_{ij}=0$, then the weights between nodes are determined only by node locations.

\begin{figure}[htbp]
\begin{center}
\includegraphics[width=.75\textwidth]{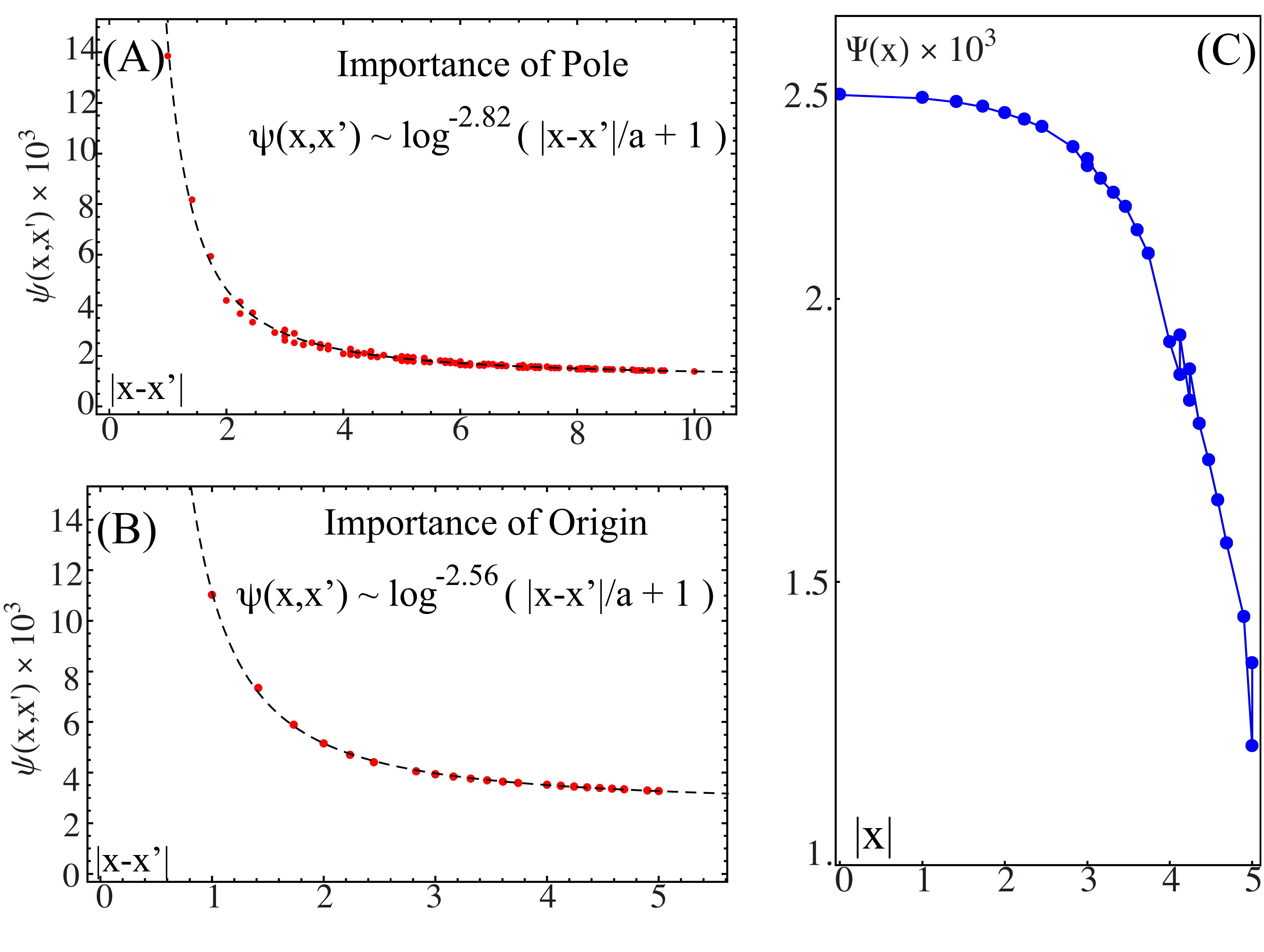}
\caption{(A) shows the importance that a node at $\xv$ assigns to a node at one of the poles of the sphere $\xv=(R,0,0)$ for all nodes in a sphere of radius $R=5a$ with a weight $u(\xv,\xv')=e^{-|\xv-\xv'|/a}$ and no long-range connections.  The importance assigned by a node at $\xv'$ towards the reference point at $\xv$ decreases with an empirically determined scaling which is not expected to be universal.  
(B) shows the importance of the origin ($\xv=(0,0,0)$) as a function of the distance $|\xv'|$ and decreases more slowly than the importance assigned to a node at the pole due to $\xv$'s central location in the network.  
(C) shows the global centrality of each node $\Psi(\xv)$ as a function of its distance from the origin.  Sharp variations for particular values of $\Psi(\xv)$ are due to nodes with a differing local connectivity but an identical distance from the origin (e.g. $|\xv|=5a$ for $\xv=(0,0,5)$ and for $\xv=(0,4,3)$).  The overall decrease in $\Psi(\xv)$ as a function of $|\xv|$ is readily apparent, with the origin being most important as expected.}  \label{ImportanceSphere.fig}
\end{center}
\end{figure}

Replacing discrete equations with continuum models has been of use in a wide variety of systems\cite{newmanContinuum}, and for networks with large $N^*$, we can establish a continuum limit of the definition of the GENs in Eq. 1 of the main text.  
In SI sec. A, we develop a linearized continuum model for the approximate pairwise importance $\phi(\xv,\xv')$ and describe a more complete (and complex) nonlinear continuum model in SI sec. D.  An inhomogeneous Fredholm equation is found for the purely geometric network (where the number of long-range network nodes $N=0$ or equivalently $w_{ij}=0$), with
\begin{eqnarray}
\phi(\xv,\xv')=\frac{u^2(\xv,\xv')}{U(\xv')}+\frac{1}{U(\xv')}\int d\zv\  u(\xv',\zv)\phi(\xv,\zv)\bigg(\rho(\zv)-\delta(\xv-\zv)\bigg),\label{continuumNoNet}
\end{eqnarray}
where $\rho(\zv)$ is the local density of material nodes at $\zv$ and $U(\xv)=\int d\zv\  \rho(\zv) u(\xv,\zv)$ is the strength the material node at $\xv$.  The delta function removes the (undefined) factor of $\phi(\xv,\xv)$ (necessary because of the discrete requirement that $E_{ii}=0$), and is equivalent to the removal of a self-energy.  Note that if $u(\xv,\xv')\ne$ constant, the linearized importance will yield a non-trivial result ($\phi(\xv,\xv')\ne$ constant).  It is straightforward to show that the solution to the Eq. \ref{continuumNoNet} is $\phi(\xv,\xv')=\bar\phi(\xv,\xv')-\bar\phi(\xv,\xv)$, where 
$U(\xv')\bar\phi(\xv,\xv')=u^2(\xv,\xv')+\int d\zv\rho(\zv)u(\xv',\zv)\bar\phi(\xv,\zv)$ is the importance without removal of the self-energy term.  There are a wide range of methods to determine $\bar\phi(\xv,\xv')$ analytically\cite{PolyaninBook,TricomiBook}, but determining an exact expression is a non-trivial task even for simple geometries (see SI Sec C).  

\begin{figure}[htbp]
\begin{center}
\includegraphics[width=.75\textwidth]{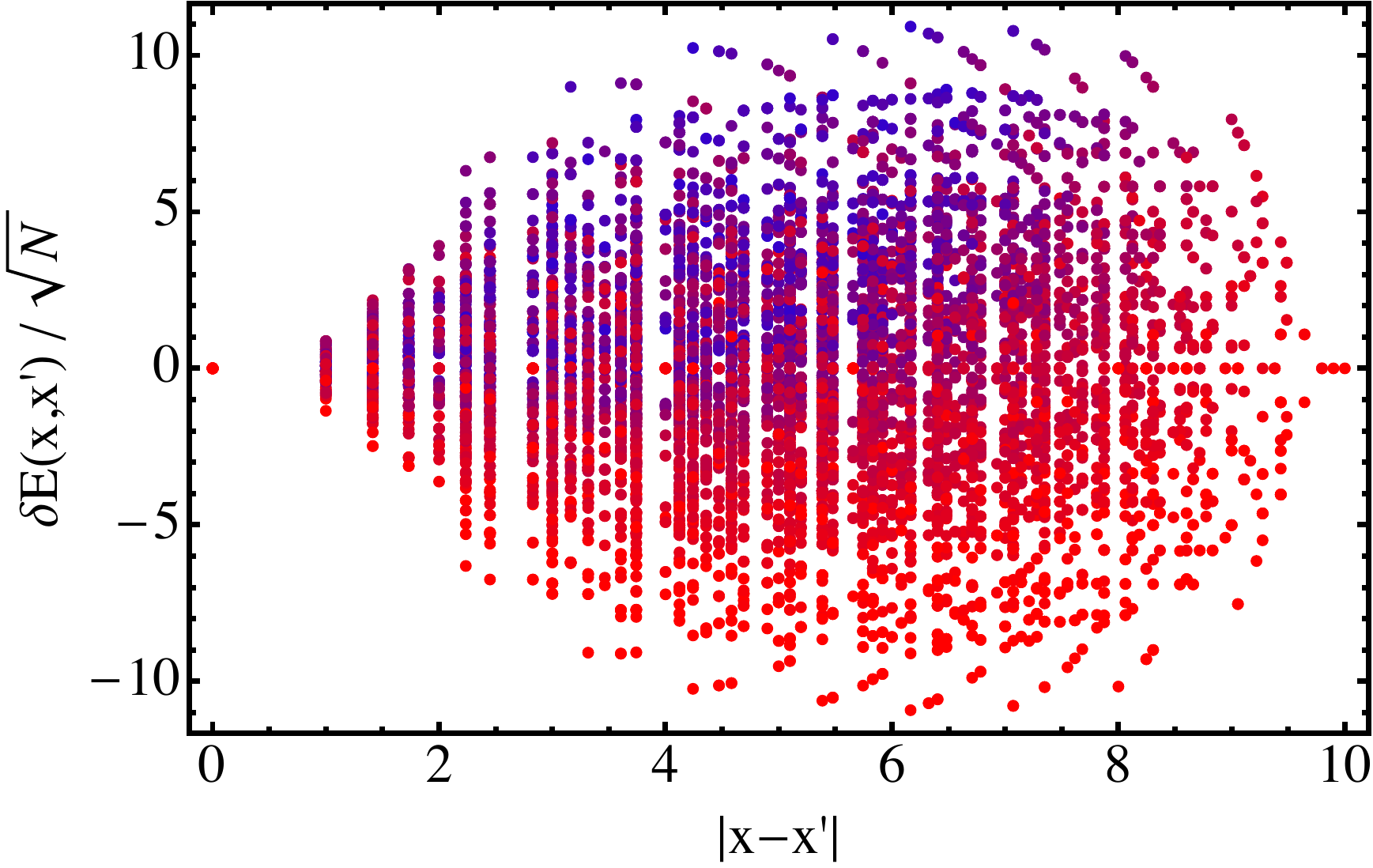}
\caption{The asymmetry $\delta E(\xv,\xv')=E(\xv,\xv')-E(\xv',\xv)$ for a sphere of radius $R=5a$ (with no long-range interactions) as a function of the distance between the nodes in the absence of any long range connections.  Deeper red points correspond to $\xv'$ near the origin and blue points correspond to $\xv'$ near the boundary of the sphere, so nodes on the periphery feel the central nodes are more important than vice versa.  
The relationship between $\delta E_{ij}$ and $\delta \tau_{ij}$ in Fig. 2 of the main text suggests that random walks reach the center of the sphere more readily than a specified point on the surface.}  
\label{UniformSphere.fig}
\end{center}
\end{figure}

If we combine short ranged geometrical interactions with long range, geometry-independent edges, the expression for the GENs and the linearized importance $\phi$ becomes more complex.  Eq. \ref{continuumNoNet} is valid only in the absence of long range connections, but we show in SI Sec. B that the linearized can still be written in the simple form 
\begin{eqnarray}
\phi(\xv,\xv')=\phi_0(\xv,\xv')+\int_\Omega d\zv K_1(\xv,\xv',\zv)\phi(\xv,\zv)\label{simpleEq},
\end{eqnarray}
with the complicated functional forms of $\phi_0(\xv,\xv')$ the kernel $K_1(\xv,\xv',\zv)$ explicitly given in SI Sec B.   While Eq. \ref{simpleEq} is linear (and therefore analytically tractable), all methods for determining $\phi(\xv,\xv')$ will require integration of functions involving $\frac{1}{U(\zv)}$, which we show in SI Sec. C is difficult even in simple domains and for simple interactions $u(\xv,\xv')$.  Eq. \ref{continuumNoNet} and \ref{simpleEq} can still be solved numerically over finite domains using well known methods\cite{PolyaninBook,TricomiBook} in cases where exact results can not be obtained.  The global importance, $\Phi(\xv)$, is straightforwardly computed from Eq. \ref{simpleEq} using $\Phi(\xv)=\int d\zv \rho(\zv)\phi(\xv,\zv)$ (assuming $u(\xv,\xv')>0$ for all $|\xv-\xv'|<\infty)$.  

While Eq. \ref{simpleEq}, describing the continuum linearized importance, is analytically approachable, the full nonlinear theory for the continuum system is entirely intractable.  Because the undesirable aspects of the linearized importance shown in Fig. \ref{LinearFailure.fig} are expected to persist even in the continuum limit, we expect the nonlinear theory to be required to yield meaningful estimates of personalized closeness or global centrality.   To understand the effect of the material interactions, we determine the GENs for a sphere of radius $R$ formed from a lattice of equally spaced material nodes with an inter-node spacing $a$ (i.e. with a constant node density $\rho$).  It is straightforward to simply solve for the GENs in Eq. 1 of the main text numerically for any geometry, which equivalent to solving the integral equation via quadrature\cite{PolyaninBook}.  For $R=5a$, the $N+N^*=515$ nodes in the network are connected with a strength $u(\xv,\xv')=e^{-\lambda |\xv-\xv'|/a}$ (with $\lambda=1$ chosen here).  For a purely material sphere (where the interaction strength is determined entirely by $u(\xv,\xv')$ and the number of network nodes is $N=0$), the closeness between two points is determined primarily by their relative distance, as shown in Fig. \ref{ImportanceSphere.fig}(A-B).  While the expected decrease in the importance of $\xv$ from the perspective of another node at $\xv'$ for increasing $|\xv-\xv'|$ is observed, the qualitative behavior of the importance depends strongly on the locations of both points, with the empirically derived scaling of $\psi(\xv,\xv')\sim \log^{\beta(\xv)}(|\xv-\xv'|/a+1$)+const.  We also find that nodes towards the center of the sphere have a higher global importance $\Phi(\xv)$ than those towards the boundary (Fig. \ref{ImportanceSphere.fig}(C)), due to the greater number of paths towards the center than along the surface.  We likewise expect the asymmetry in the closeness, $\delta E(\xv,\xv')=E(\xv,\xv')-E(\xv',\xv)$, is skewed such that more external nodes feel closer to internal nodes than vice versa, due to the greater overall importance of the central nodes.  This is confirmed in Fig. \ref{UniformSphere.fig}, where $\delta E(\xv,\xv')<0$ is far more likely to be seen at the boundary of the sphere, regardless of the distance between the nodes.  Coupled with Fig. 2 of the main text we expect that $\delta \tau(\xv,\xv')\sim -\delta E(\xv,\xv')\times \sqrt{\alpha N}$, for some topology-dependent $\alpha$, will accurately predict the asymmetry in the mean first passage time of a random walk between nodes in the sphere.

\begin{figure}[htbp]
\begin{center}
\includegraphics[width=\textwidth]{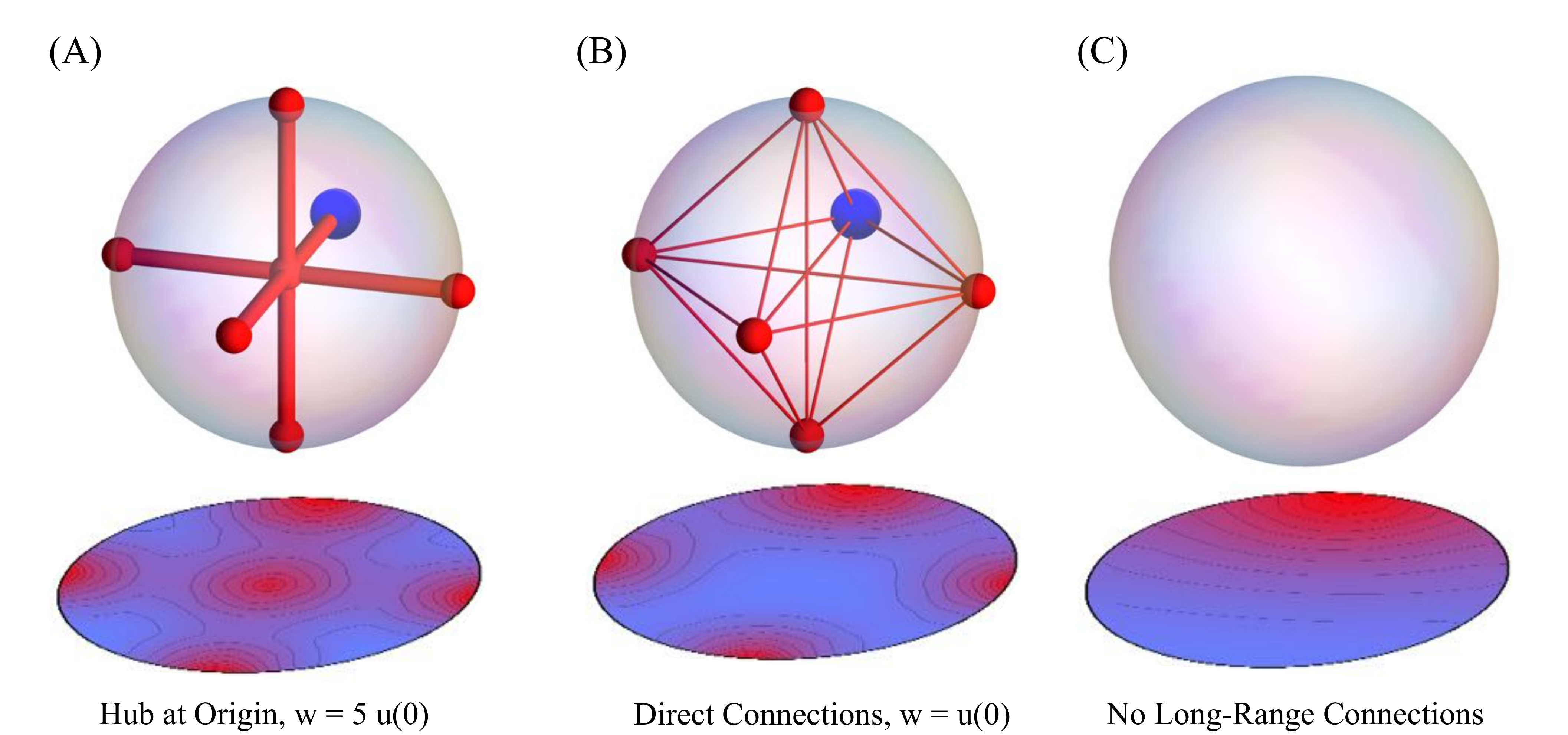}
\caption{The importance from a node located in the $x-y$ plane felt towards the node at $\xv=(R,0,0)$.  A schematic of the network topology is shown on top, and the importance $\psi(\xv,\xv')$ for all points in the $x-y$ plane are shown below, with red corresponding to a higher importance.  (A) shows the effect of long-range connections between the poles and the center of the sphere (with each long-range edge having weight $w_{ij}=5$) for all edges relative to $\xv=(R,0,0)$ (marked with the blue sphere).  While each pole feels somewhat closer to nearby nodes than it does to distant nodes, it is still clear that the poles are strongly connected.  (B) shows a network with the same strength of all nodes but the origin and a differing topology of direct connections between poles with weight $w_{ij}=1$.  Surprisingly, the importance assigned between poles is lower in (B) than (A) (i.e. stronger indirect connections produce a greater pole-to-pole importance than weaker direct connections), and the range of importance due to the distance-dependence connections is increased.  (C) shows the material-only case with $N=0$ (shown in Fig. \ref{ImportanceSphere.fig} (A) as well). } \label{SphereWithNetwork.fig}
\end{center}
\end{figure}

The influence of long-range network topology greatly increases the complexity of the problem and obscures our understanding of the closeness between nodes in the sphere.  To study the influence of the topology of the long range connections, we consider two very simple networks embedded in our sphere:  long range connections between the poles of the sphere passing through a central hub, and direct connections between the poles (pictured in Fig. \ref{SphereWithNetwork.fig}).  In Fig. \ref{SphereWithNetwork.fig}(A), we show the network nodes at the poles are indirectly linked to each other through strong connections to the center of the sphere of strength $w_{ij}=5$  (so $W=\frac{1}{2}\sum_{ij}w_{ij}=30)$.  Also shown is a cross-section of the sphere, showing the closeness between points to one of the poles (marked with the blue dot in the sphere above).  The total strength of the distance-dependent interactions is $U(\xv)=\sum_{l} u(\xv,\xv_l)\approx 7.6$ at the poles, so the contribution of the material nodes is comparable to that of the long-range connections.  Unsurprisingly, the center of the sphere feels each of the poles are important (the central red region in the cross-section in Fig. \ref{SphereWithNetwork.fig}(A)), as do each of the other network-node poles.  We compare the central-hub topology in Fig. \ref{SphereWithNetwork.fig}(A) to that of direct connections between poles, depicted in Fig. \ref{SphereWithNetwork.fig}(B).  To keep the strength of the poles constant, the long range connections all have weight $w_{ij}=1$ (but $W=15$).  Despite the constant total strength of the poles, the increased number of edges per pole decreases the importance each pole assigns to the other, reflected in the fact that the pole-to-pole importance, $\psi(\xv,\xv')$, in (B) is reduced by about 5\% relative to (A).  Indirect connections between the poles can therefore lead to a greater importance between them than direct connections due to the effect of short-ranged interactions with material nodes.   The decay length of the importance assigned by material nodes to the network nodes is also increased for the direct connections in (B) relative to the range in (A) (not shown), suggesting that the reduced importance between network nodes is due to an increased importance of the material nodes.  In situations where finite resources are shared (e.g. a random walk or diffusion of information), direct connections to a central hub will more easily allow the resource to be targeted to specific sites in comparison to multiple direct connections.

\section{Conclusions}

In this paper, we have shown the utility of the Generalized Erd\H{o}s Numbers in a variety of contexts on both a local and a global level.  The  asymmetry of the GENs is highly correlated to the asymmetry in the MFPT, another asymmetric quantity of use in network science.  The ability of the GENs to generate a global measure of closeness between nodes in a network has been shown in the context of defining both a global and personalized measure of the importance or centrality.  Our global Erd\H{o}s centrality is consistent with other well known centrality measures with the added benefit of a greater heterogeneity for nodes of equal degree in many cases (see Fig. {ErdosCentrality.fig}).  The personalized importance $\psi_{ij}$ from which the Erd\H{o}s centrality is derived is shown to be useful in other contexts, with a specific example studying the influence specific locations of a sphere can have on other locations. Having demonstrated the usefulness of the GENs in many contexts here, future work will apply these methods to concrete problems involving network growth or dynamics on networks in more concrete cases of physical or biological importance.

Acknowledgements:  We thank O. Peleg and G. Strang for useful comments on the manuscript.  {\bf{grant info}}

%

\appendix

 \section{Formulation of the linearized continuum model}
 \label{ContinuumStrength}

In order to illustrate the emergence of both the node density and the delta function constraints in the continuum limit (Eq. \ref{continuumNoNet} of the main text), it is useful to begin by calculating $U(\xv)$ in the case where network nodes are not included (or only between material nodes).  We presume that there are a sufficiently large number of material nodes $N^*$ so that we can replace the sums over all nodes (such as in Eq. \ref{GENdef} of the main text) with an integral.  The interaction between material the nodes $i$ and $j$ is $u(\xv_i,\xv_j)$.  For the discrete system, the total strength of the material node $j$ is given by
\begin{eqnarray}
U_j=\sum_{i\ne j}u(\xv_i,\xv_j)=\sum_{i}u(\xv_i,\xv_j)-u(\xv_j,\xv_j),
\end{eqnarray}
where the second term imposes the constraint of no self-interactions (no loops) in the network.  This constraint can be avoided by allowing self-loops and will have no impact on the continuum model for the linearized importance, but such a change would have an minor impact on the non-linear theory.  Because the GENs were originally developed for networks without self-interactions, we retain that assumption in this paper.  If each material node were on a lattice with spacing $a$, the volume excluded by each node would be $a^d$.  We could then write
\begin{eqnarray}
U_i=\sum_j \frac{a^d u(\xv_i,\xv_j)}{a^d}-u(\xv_i,\xv_i)\to \int_\Omega d\zv \ \rho u(\zv,\xv_i)\bigg[1-\delta(\xv-\zv)\bigg]\equiv U(\xv_j)-u(\xv_j,\xv_j)\label{continuumStrength}
\end{eqnarray}
in the limit of $N^*\to \infty$ and $a\to 0$, where $\rho=1/a^d$ is the constant number density of the nodes at the location $\zv$ and with the constraint $\rho V=N^*$, where $V$ is the volume of the domain and  $U(\xv)$ is the total strength with the integral over all space (i.e. without neglecting the self-loop).  We have kept the continuum version of $U(\xv)$ to be the integral over all space for convenience, but must remember to remove the contribution from the self loop as in Eq. \ref{continuumStrength}.  For material nodes that are not on a lattice, we can replace the effective volume $a^d$ with the volume excluded by the node $j$ $v(\xv_i)$, and defining the variable density, $\rho(\xv)=1/v(\xv)$, produces the strength 
\begin{eqnarray}
U(\xv)=\sum_j \frac{v(\xv_j) u(\xv_i,\xv_j)}{v(\xv_j)}\to\int d\zv \rho(\zv)u(\zv,\xv)
\end{eqnarray}
where again $U_j\to U(\xv_j)-u(\xv_j,\xv_j)$.  A similar limit will be found for any other sum that must avoid self-loops in the continuum limit, such as in Eq. \ref{GENdef} of the main text.

\section{Appendix:  Including long-range connections in the continuum model}
\label{ContinuumLongRangeAppendix}
Labeling $\phi^{(n\to m)}(\xv,\yv)$ as the importance a network node at $\yv$ (connected to a long range edge) assigns to a material node at $\xv$ (with no long range connection), the linearized importance between material nodes reduces to the coupled equations
\begin{eqnarray}
U(\xv')\phi(\xv,\xv')&=&u^2(\xv,\xv')+\int d\zv\ u(\xv',\zv)\phi(\xv,\zv)\bigg[\rho(\zv)-\delta(\xv-\zv)\bigg]\nonumber\\
&&+\sum_k u(\xv',\yv_k)\blp \phi^{(n\to m)}(\xv,\yv_k)-\phi(\xv,\yv_k)\brp\label{Coupled1}
\end{eqnarray}
and
\begin{eqnarray}
\blp U(\yv_l)+W_k\brp\phi^{(n\to m)}(\xv,\yv_k)&=&u^2(\xv,\yv_k)+\int d\zv\ u(\yv_k,\zv)\phi(\xv,\zv)\bigg[\rho(\zv)-\delta(\xv-\zv)\bigg]\\
&&+\sum_l [w_{kl}+u(\yv_k,\yv_l)] \phi^{(n\to m)}(\xv,\yv_l)-u(\yv_k,\yv_l)\phi(\xv,\yv_l)\nonumber,
\label{Coupled2}
\end{eqnarray}
where Eq. \ref{Coupled1} describes the importance between material nodes, and Eq. \ref{Coupled2} describes the importance from the $N$ network nodes to the material nodes.  Eq. \ref{Coupled1} is identical to Eq. \ref{continuumNoNet} of the main text with a set of additional $\delta$-function constraints at the location of the network nodes (where long-range connections contribute).  Note that if $w_{ij}=0$, the substitution of $\phi^{(n\to m)}(\xv,\yv_l)=\phi(\xv,\yv_l)$ into Eq. \ref{Coupled2} recovers Eq. \ref{Coupled1}, which in turn reduces to Eq. \ref{continuumNoNet} of the main text as expected.  The discrete nature of the network nodes allows a direct solution in terms of $\phi(\xv,\xv')$, and it is convenient to define $\Mv=(\Lv_u+\Lv_0)^{-1}$, with $(\Lv_0)_{ij}=W_{i}\delta_{ij}-w_{ij}$ the un-normalized graph Laplacian of the network nodes (in the absence of any material nodes) and $(\Lv_d)_{ij}=U(\yv_i)\delta_{ij}-u(\yv_i,\yv_j)$ incorporates the effects of the material nodes on the strength of the network nodes.  It is important to note that $\Lv_d+\Lv_0$ is {\em{not}} the graph Laplacian of the entire network, and can be inverted so long as $u(\zv,\zv')>0$.  It is straightforward to derive the solution for $\phi^{(n\to m)}$ in terms of $\phi$, with
\begin{eqnarray}
[(\Lv_u+\Lv_0)\bm{\phi}^{(n\to m)}(\xv)]_k=u^2(\xv,\yv_l)+\int d\zv\phi(\xv,\zv)u(\yv_l,\zv)\bigg[\rho(\zv)-\delta(\xv-\zv)\bigg]\label{matrixEq}\\
-\sum_m\phi(\xv,\yv_m)u(\yv_l,\yv_m)\nonumber,
\end{eqnarray}
where $[\bm{\phi}^{(n\to m)}(\xv)]_k=\phi^{(n\to m)}(\xv,\xv_k)$.  Eq. \ref{matrixEq} can be solved directly, and we find and substitution  into Eq. \ref{Coupled1} yields
\begin{eqnarray}
U(\xv')\phi(\xv,\xv')&=&u^2(\xv,\xv')+\sum_{kl}u(\xv',\yv_k) \Mv_{kl}u^2(\xv,\yv_l)-\sum_lu(\xv',\yv_l)\phi(\xv,\yv_l)\nonumber\\
&&+\int d\zv\phi(\xv,\zv)[\rho(\zv)-\delta(\xv-\zv)]\bigg[u(\xv',\zv)+\sum_{kl}u(\xv',\yv_k) \Mv_{kl}u(\yv_l,\zv)\bigg]\nonumber\\
&&-\sum_{kl}u(\xv',\yv_k) \Mv_{kl}\sum_m\phi(\xv,\yv_m)u(\yv_l,\yv_m)\label{LinearFinal}
\end{eqnarray} 
where $\Mv=(\Lv_u+\Lv_0)^{-1}$.  Eq. \ref{LinearFinal} is linear in $\phi(\xv,\xv')$, meaning that it can be solved exactly knowing the locations of and interactions between network nodes.  In particular, we can write
\begin{eqnarray}
\phi(\xv,\xv')=\phi_0(\xv,\xv')+\int_\Omega d\zv K_1(\xv,\xv',\zv)\phi(\xv,\zv)\label{simpleEq2}
\end{eqnarray}
with
\begin{eqnarray}
\phi_0(\xv,\xv')=\frac{u^2(\xv,\xv')}{U(\xv')}+\frac{1}{U(\xv')}\sum_{kl}u(\xv',\yv_k) \Mv_{kl}u^2(\xv,\yv_l)
\end{eqnarray}
is the direct connection term, renormalized to include the effects of the long-distance connection, and the kernel becomes 
\begin{eqnarray}
K_1(\xv,\xv',\zv)=\bigg[1-\frac{\delta(\xv-\zv)}{\rho(\xv)}-\sum_m\frac{\delta(\yv_m-\zv)}{\rho(\yv_m)}\bigg]K_1^{(b)}(\xv',\zv)
\end{eqnarray}
with 
\begin{eqnarray}
K_1^{(b)}(\xv',\zv)=\frac{\rho(\zv)}{U(\xv')}\bigg[u(\xv',\zv)+\sum_{kl}u(\xv',\yv_k)\Mv_{kl}u(\yv_l,\zv)\bigg]
\end{eqnarray}
 is a `bare' kernel (in the absence of any delta function constraints).  It is worthwhile to note that the the dominant contribution to $\phi(\xv,\xv')$ comes from $\phi_0(\xv,\xv')$ in Eq. \ref{simpleEq2} (with higher order contributions coming from integrals weighted by $u$).  If long-range interactions are expected to be perturbative and local interactions are very short-ranged,  $\phi_0(\xv,\xv')$ may be a sufficient approximation to avoid the complexity of an exact evaluation of $\phi(\xv,\xv')$.  However, the undesirable behavior of the linearized importance, pictured in Fig. \ref{LinearFailure.fig} of the main text, suggests that this further-approximated value of $\phi(\xv,\xv')$ should be used with caution.

The most straightforward approach to solving Eq. \ref{simpleEq} of the main text is the development of a Neumann series, and it is simple to show that defining $\phi(\xv,\xv')=\sum_{n=0}^\infty \phi_n(\xv,\xv')$ with $\phi_n(\xv,\xv')=\int d\zv K_n(\xv,\xv',\zv)\phi_0(\xv,\zv)$ for $n\ge 1$ and $K_n(\xv,\xv',\zv)=\int d\zv' K_1(\xv,\xv',\zv')K_{n-1}(\xv,\zv',\zv)$.  The delta function constraints make this form of the kernel difficult to work with, but it is possible to reduce $K_n(\xv,\xv',\zv)$ in terms of the more easily computed $K_n^{(b)}(\xv',\zv)=\int d\zv K_1^{(b)}(\xv',\zv')K_{n-1}^{(b)}(\zv',\zv)$.  It is not difficult to show that for a given location $\xv$ and known $\{\yv_l\}$ that the full kernel has the form 
\begin{eqnarray}
K_n(\xv,\xv',\zv)&=& \blp 1-\frac{\delta(\zv-\xv)}{\rho(\xv)}-\sum_l\frac{\delta(\zv-\yv_l)}{\rho(\yv_l)}\brp\blp K_n^{(b)}(\xv',\zv)\\
&&+\sum_{m=1}^{n-1}\bigg[b_{m}K_m^{(b)}(\xv',\zv)+c_{m}K_m^{(b)}(\xv',\zv)+\sum_l d_{ml}K_m^{(b)}(\xv',\yv_l)\bigg]\brp\nonumber,
\end{eqnarray}
with large and unwieldy recursive relations defining the coefficients $b_m$, $c_m$ and $d_{ml}$, each of which depend on $\xv$ and $\{\yv_l\}$.  Importantly, though, the coefficients do not depend on $\zv$, the variable of integration in the definition of $\phi_n(\xv,\xv')$.  It is therefore in principle possible to determine $\phi(\xv,\xv')$ exactly with knowledge of $K_n^{(b)}(\xv,\xv')$, which can be determined using standard methods.  However, as noted in SI Sec. \ref{Ucalculation}, integrals over $1/U(\xv')$ are expected to be very unwieldy, and numerical work is likely required to determine the propagator.

 \section{Determining $U(x)$ for a simple interaction}
 \label{Ucalculation}
 
 As long as our domain is finite, material nodes on the boundary of the domain will be less connected than material points in the interior.  This will be reflected in a non-constant $U(\xv)$, which gives rise to a variable importance of the material nodes.  Computing $U(\xv)$ is not necessarily trivial.  If we suppose that $u(\xv,\xv')=u(|\xv-\xv'|)$ and that the domain is a sphere of radius $R$ with a constant node density $\rho(\xv)=V^{-1}$, we find
 \begin{eqnarray}
 U(\xv)=\frac{1}{V}\int_\Omega d^3\xv' u(|\xv-\xv'|)=\frac{1}{V}\int\frac{d^3\kv}{(2\pi)^3}e^{i\kv\cdot\xv}\hat u(|\kv|)\int_\Omega d^3\yv e^{-i\kv\cdot\yv}\\
 =\frac{1}{V}\int_0^\infty \frac{dk}{2\pi} \frac{4\hat u(k)}{k^2x}\blp \sin(kR)-kR\cos(kR)\brp\sin(k x),
 \end{eqnarray}
 where $\hat u(\kv)$ is the Fourier Transform of $u(\xv)$.  As a specific example, if $u(|\xv-\yv|)=e^{-\l|\xv-\yv|}$, with $\hat u(k)= 8\pi \l/(k^2+\l^2)^2$ in three dimensions, we find
 \begin{eqnarray}
 U(x)=\frac{6}{R^3\l^3}+\frac{3 e^{-\l R}}{R^3\l^4 x}\blp \cosh(\l x)[\l x+\l^2 x R]-\sinh(\l r)[\l^2 R^2-3\l R-3]\brp\label{threeDweight},
 \end{eqnarray}
yielding an expression for the total strength at the material node $x$ expressed in terms of elementary functions only.  Other forms of the interaction strength (for example, a gaussian) tend to produce more complex expressions, and we choose to use only an exponential decay for the interaction strength between material nodes.  Regardless of the (relative) simplicity of Eq. \ref{threeDweight}, terms involving integrals of $1/U(\xv')$ will be difficult to evaluate exactly (as the reader may readily verify), meaning that the Neumann series\cite{PolyaninBook,TricomiBook} described in appendix \ref{ContinuumLongRangeAppendix} is not easily computed analytically.

 \section{The Nonlinear continuum model}
 \label{NonlinearContinuum}

The development of the continuum model for the nonlinear GENs follows the construction in Appendix \ref{ContinuumStrength} in a straightforward fashion.  The discrete case of the GENs constrain $E_{ii}=0$, but in the continuum limit we will have a self-energy contribution (similar to that of the self-interaction due to $u(\xv_i,\xv_i)\ne 0$ above). The $\delta$-function constraints for the nonlinear GENs is most easily developed by writing
\begin{eqnarray}
\frac{U_j}{E(\xv_i,\xv_j)}&=&u^2(\xv_i,\xv_j)+\sum_l \frac{u^2(\xv_j,\xv_l)}{E(\xv_i,\xv_l)u(\xv_j,\xv_l)+1}\nonumber\\
\qquad\qquad\qquad&&-\frac{u^2(\xv_i,\xv_j)}{E(\xv_i,\xv_i)u(\xv_i,\xv_j)+1}-\frac{u^2(\xv_j,\xv_j)}{E(\xv_i,\xv_j)u(\xv_j,\xv_j)+1},
\end{eqnarray}
where the first term on the right hand side correctly handles the direct connection between nodes $i$ and $j$; the second term is the sum over all nodes (including $i$ and $j$); the third term removes the contribution due to a non-zero $E(\xv_i,\xv_i)$; and the fourth term removes the contribution due to the direct connection between the nodes at $\xv_i$ and $xv_j$.  It is not difficult to see that in the continuum limit, this expression becomes
\begin{eqnarray}
\frac{U(\xv')-u(\xv,\xv')}{E(\xv,\xv')}+\frac{u(\xv',\xv')}{E(\xv,\xv')u(\xv,\xv')+1}&=&u^2(\xv,\xv')+\int d\zv\ \rho(\zv)\frac{u^2(\xv',\zv)}{E(\xv,\zv)u(\xv',\zv)+1}\nonumber\\
\qquad\qquad\qquad&&-\frac{u^2(\xv,\xv')}{E(\xv,\xv)u(\xv,\xv')+1}\label{Nonlinear1}
\end{eqnarray}
after recalling that $U_j=U(\xv_j)-u(\xv_j,\xv_j)$.  The second term on the left hand side of the equation is due to the removal of the self-loop.  It is clear that Eq. \ref{Nonlinear1} can not be expressed in terms of any standard integral equation, and the nonlinearity guarantees that analytic work is unlikely to be fruitful without further approximation.  If we assume that $E(\xv,\xv')\gg u^{-1}(\xv,\xv')$ for all $\xv$ and $\xv'$ and define $\phi(\xv,\xv')\sim E^{-1}(\xv,\xv')$, it is straightforward to see that
\begin{eqnarray}
U(\xv')\phi(\xv,\xv')=u^2(\xv,\xv')+\int d\zv\ \phi(\xv,\zv)u(\xv',\zv)\blp \rho(\zv)-\delta(\xv-\zv)\brp,
\end{eqnarray}
which is Eq. \ref{continuumNoNet} of the main text.  The fact that Eq. \ref{continuumNoNet} of the main text is the well known inhomogeneous Fredholm Equation\cite{PolyaninBook,TricomiBook} means that analytic work is possible using the linearized importance.  The linearized importance was discussed in more detail in Appendix \ref{ContinuumLongRangeAppendix}.

In the continuum model, the long range connections between nodes at the specified points $\{\yv_l\}$ must be enforced by delta-function constraints (since the material nodes an infinitesimal distance away from $\yv_j$ will not have a long-range connection), much like the self-loops are removed.  The closeness between material nodes, $E(\xv_i,\xv_j)$, is still expected to be a continuous function, but we must also account for the long-range connections between network nodes.  Because Eq. \ref{GENdef} of the main text determines the closeness felt by a network node $\yv_j$ towards a material node $\xv_i$ in terms of the direct connection to other material nodes $\{\yv_l\}$ (with strength $w_{jl}$) and the closeness felt by $\yv_l$ towards $\xv_i$.  When computing the closeness felt by any node towards a material node, we do {\em{not}} need to compute the closeness felt by any node towards a network node.  It suffices to compute the material-material closeness $E(\xv_i,\xv_j)$ as well as the network-material closeness $E^{(n\to m)}(\xv_i,\yv_j)$.  In the discrete case, this labeling is irrelevant, but it is essential in the continuous case.  We then write
\begin{eqnarray}
\frac{U(\xv_j)-u(\xv_j,\xv_j)}{E(\xv_i,\xv_j)}&=&u^2(\xv_i,\xv_j)+\sum_{\substack{l\ne i,j\\ \xv_l\not\in\{\yv_l\} }}\frac{u^2(\xv_j,\xv_l)}{E(\xv_i,\xv_l)u(\xv_j,\xv_l)+1}\nonumber\\
&&\qquad\qquad\qquad\qquad+\sum_{\yv_l} \frac{u^2(\xv_j,\yv_l)}{E^{(n\to m)}(\xv_i,\xv_l)u(\xv_j,\xv_l)+1}\label{discreteNonlinear1}
\end{eqnarray}
and
\begin{eqnarray}
\frac{U(\yv_i)+W_j-u(\yv_j,\yv_j)}{E^{(n\to m)}(\xv_i,\yv_j)}&=&u^2(\xv_i,\yv_j)+\sum_{\substack{l\ne i,j\\ \xv_l\not\in\{\yv_l\} }}\frac{u^2(\yv_j,\xv_l)}{E(\xv_i,\xv_l)u(\yv_j,\xv_l)+1}\nonumber\\
&&\qquad\qquad\qquad\qquad+\sum_{\yv_l\ne \yv_j} \frac{u^2(\yv_j,\yv_l)}{E^{(n\to m)}(\xv_i,\yv_l)u(\yv_j,\yv_l)+1}\label{discreteNonlinear2}.
\end{eqnarray}
While we have clearly delineated the distinction between $E(\xv,\xv')$ and $E^{(n\to m)}(\xv,\yv')$, there is no approximation here and Eqs. \ref{discreteNonlinear1} and \ref{discreteNonlinear2} are identical to Eq. \ref{GENdef} of the main text.  In the continuum limit, only the sum over the material nodes can be converted to an integral, as the number of network nodes $N$ remain fixed.  We  can then write
\begin{eqnarray}
\frac{U(\xv')-u(\xv',\xv')}{E(\xv,\xv')}+\frac{u^2(\xv',\xv')}{E(\xv,\xv')u(\xv',\xv')+1}=u^2(\xv,\xv')+\sum_l\frac{u^2(\xv',\yv_l)}{E^{n\to m}(\xv,\yv_l)u(\xv',\yv_l)+1}\\
+\int d\zv \frac{u^2(\xv',\zv)}{E(\xv,\zv)u(\xv',\zv)+1}\blp \rho(\zv)-\delta(\xv-\zv)-\sum_l \delta(\yv_l-\zv)\brp\nonumber
\end{eqnarray}
and
\begin{eqnarray}
\frac{U(\yv_j)-u(\yv_j,\yv_j)+W_j}{E^{(n\to m)}(\xv,\yv_j)}+\frac{u^2(\yv_j,\yv_j)}{E^{(n\to m)}(\xv,\yv_j)u(\yv_j,\yv_j)+1}=u^2(\xv,\yv_j)\\
+\sum_l\frac{[u(\yv_j,\yv_l)+w_{jl}]^2}{E^{n\to m}(\xv,\yv_l)[u(\yv_j,\yv_l)+w_{jl}]+1}\nonumber\\
+\int d\zv \frac{u^2(\yv_j,\zv)}{E(\xv,\zv)u(\yv_j,\zv)+1}\blp \rho(\zv)-\delta(\xv-\zv)-\sum_l \delta(\yv_l-\zv)\brp\nonumber
\end{eqnarray}
The full nonlinear version of the continuum limit is again clearly not analytically approachable, but the linearized version can be reduced to a single Fredholm equation, as is discussed further in Appendix \ref{ContinuumLongRangeAppendix}.  If desired, the closeness felt by a material node towards a network node can be similarly computed (after defining a material-to-network closeness $E^{(m\to n)}(\yv_i,\xv)$), which we do not explicitly compute here.

\section{Convergence of the Neumann Series}

It is important to note that if $u(\xv,\xv')=u(|\xv-\xv'|)$, which is a natural choice for the form of the location dependence of the short range interactions, it is unlikely that the Neumann series will converge in an unbounded domain.  The convergence of the series $\sum_n \phi_n^{(b)}(\xv,\xv')$ is dependent on the fact that the $l_2$ norm of the kernel is bounded over the domain.  For constant $\rho$ and in an unbounded domain $U(\xv')=U$ is a constant for all $\xv$ (since every node is connected in the same way to the other nodes in the infinite space) and it is straightforward to show that the $l_2$ norm is
\begin{eqnarray}
\int d\xv' d\zv \rho(\xv')\rho(\zv)\bigg[K_1^{(b)}(\xv',\zv)\bigg]^2=\frac{\rho}{U^2}\int d\zv d\xv' u(|\xv'-\zv|)\to \infty
\end{eqnarray}
after a change of variables to $\zv'=\zv-\xv'$ (valid due to the domain's infinite extent).  For any infinite domain, then, $\phi^{(b)}(\xv,\xv')$ is undefined.  With the $\delta$-function constraints included, there will not be a divergence in $\phi$, but the sum $\phi(\xv,\xv')=\sum_n\phi_n(\xv,\xv')$ will be weakly converging (e.g. a divergent sequence of alternating sign), and while theoretically exact will likely not be useful if numerical work is required.  These issues may all be avoided if the domain is finite with volume $V$, at the cost of a more complex form for $U(\xv)$ (described in SI Sec. \ref{Ucalculation}).

\end{document}